\documentclass[aps,twocolumn,pra,superscriptaddress,footinbib]{revtex4-1}
\usepackage{amsmath,amssymb,bm}
\usepackage{graphicx}
\usepackage{epstopdf}
\usepackage{latexsym}
\usepackage[normalem]{ulem} 
\usepackage[usenames,dvipsnames]{color}
\usepackage[hidelinks]{hyperref}
\usepackage{natbib}
\usepackage{braket}
\usepackage{changepage}

\begin{document}

\title{Exploring dynamical phase transitions with cold atoms in an optical cavity}

\author{Juan A. Muniz}
\thanks{These authors contributed equally}
\affiliation{JILA, NIST, Department of Physics, University of Colorado,  Boulder, CO 80309, USA}
\author{Diego Barberena}
\thanks{These authors contributed equally}
\affiliation{JILA, NIST, Department of Physics, University of Colorado,  Boulder, CO 80309, USA}
\affiliation{Center for Theory of Quantum Matter, University of Colorado, Boulder, CO 80309, USA}
\author{Robert J. Lewis-Swan}
\thanks{These authors contributed equally}
\affiliation{JILA, NIST, Department of Physics, University of Colorado,  Boulder, CO 80309, USA}
\affiliation{Center for Theory of Quantum Matter, University of Colorado, Boulder, CO 80309, USA}
\author{Dylan J. Young}
\affiliation{JILA, NIST, Department of Physics, University of Colorado,  Boulder, CO 80309, USA}
\author{Julia R. K. Cline}
\affiliation{JILA, NIST, Department of Physics, University of Colorado,  Boulder, CO 80309, USA}
\author{Ana Maria Rey}
\affiliation{JILA, NIST, Department of Physics, University of Colorado,  Boulder, CO 80309, USA}
\affiliation{Center for Theory of Quantum Matter, University of Colorado, Boulder, CO 80309, USA}
\author{James K. Thompson}
\affiliation{JILA, NIST, Department of Physics, University of Colorado,  Boulder, CO 80309, USA}

\begin{abstract}
Interactions between atoms and light in optical cavities provide a means of investigating collective (many-body) quantum physics in controlled environments. Such ensembles of atoms in cavities have been proposed for studying collective quantum spin models, where the atomic internal levels mimic a spin degree of freedom and interact through long-range interactions tunable by changing the cavity parameters \cite{Leroux_2010,Norcia_OAT_2018,Davis_2019,Lev_2018}. 
Non-classical steady-state phases arising from the interplay between atom–light interactions and dissipation of light from the cavity have previously been investigated \cite{Baumann_2010,Hemmerich_2015,Baden2014,Ritsch_2013,Landin2018,Kroeze2018,Kroeze2019}. These systems also offer the opportunity to study dynamical phases of matter that are precluded from existence at equilibrium but can be stabilized by driving a system out of equilibrium \cite{Heyl_DPTtheory_2013,Silva_DQPT_2018,Eckstein_2009,Moore_2012,Rahul_2015}, as demonstrated by recent experiments \cite{Monroe_DPT_2017,Flaschner2018,Roos_DPT_2017,Thywissen_DPT_2018,Zhang_2017,Choi_2017}. These phases can also display universal behaviours akin to standard equilibrium phase transitions \cite{Ritsch_2013,Oberthaler_Universal_2018,Schmiedmayer_Universal_2018}. Here, we use an ensemble of about a million strontium-88 atoms in an optical cavity to simulate a collective Lipkin–Meshkov–Glick model \cite{LMG,Vidal_LMG_2007}, an iconic model in quantum magnetism, and report the observation of distinct dynamical phases of matter in this system. Our system allows us to probe the dependence of dynamical phase transitions on system size, initial state and other parameters. These observations can be linked to similar dynamical phases in related systems, including the Josephson effect in superfluid helium \cite{Backhaus1998}, or coupled atomic \cite{Oberthaler_Tunneling_2005} and solid-state polariton \cite{Abbarchi_2013} condensates. The system itself offers potential for generation of metrologically useful entangled states in optical transitions, which could permit quantum enhancement in state-of-the-art atomic clocks \cite{Campbell_2017,Ludlow_2015}.
\end{abstract}

\maketitle

\section*{Introduction}
Arrays of ultracold alkaline-earth atoms with narrow linewidth optical transitions are the basis of the most precise atomic clocks \cite{Ludlow_2015} and are also used for quantum simulation \cite{Cazalilla_2014} and quantum information processing \cite{Daley2011}. When these atoms are placed inside an optical cavity, their long-lived internal levels make them ideal to simulate non-equilibrium quantum magnetism, including models featuring long-range interactions mediated by cavity photons.

\begin{figure*}[!ht]
\includegraphics[width=6.5in]{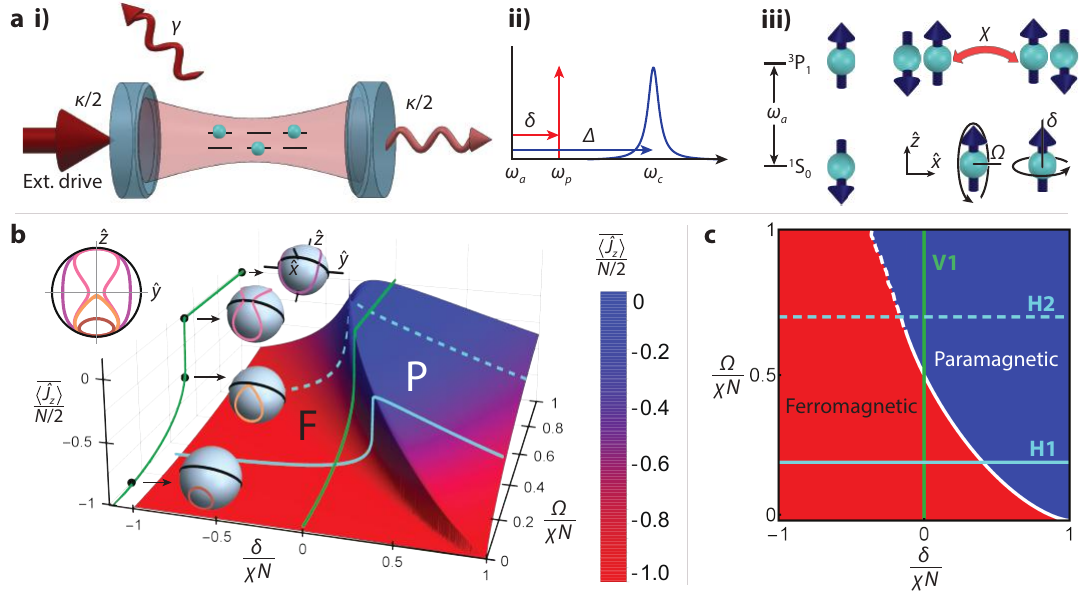}
\caption{{\bf System and dynamical phase diagram.} (a) (i) An ensemble of $^{88}$Sr atoms is trapped in a 1D optical lattice supported by an optical cavity. The atoms are coupled to a single cavity mode with a single-photon Rabi frequency 2$g$ and a resonance frequency $\omega_c$ detuned by $\Delta = \omega_c-\omega_a$ from the optical atomic transition $^1$S$_0$, $m_J=0$ ($\ket{\downarrow}$) to  $^3$P$_1$, $m_J = 0$ ($\ket{\uparrow}$) (with frequency $\omega_a$ and linewidth $\gamma$). Light leaks out of the cavity at a total rate $\kappa$. The cavity is driven externally by a laser with frequency $\omega_p$ that, if on resonance with an empty cavity, would establish a coherent state inside the cavity with average intracavity photon number $|2\Omega_p/\kappa|^2$. As shown in (ii) and (iii), for the far-detuned cavity system in consideration, the external drive generates a transverse field that drives Rabi flopping at frequency $\Omega = -2g\Omega_p/\Delta$. The external drive detuning $\delta = \omega_p - \omega_a$ establishes a longitudinal field. The detuned cavity field generates an effective spin exchange interaction of strength $\chi = -g^2/\Delta$ as shown in (iii). (b) The collective LMG model with transverse and longitudinal fields  features a second-order DPT between paramagnetic (P, blue) and ferromagnetic phases (F, red). The DPT is characterized by the long-time average of the collective magnetization $\overline{\langle \hat{J}_z \rangle}$, and its dynamics can be characterized by trajectories of the classical Bloch vector in the pseudospin Bloch sphere (see projection and associated sphere insets). For $\delta = 0$, in the paramagnetic phase the trajectories circumnavigate the Bloch sphere, whereas in the ferromagnetic phase the trajectories are trapped below the equator. (c) The two-dimensional map shows the DPT indicated by a sharp change in $\overline{\langle \hat{J}_z \rangle}$ (white solid line) for $\delta/(\chi N) \geq -1/8$. The white dashed line ($\delta/(\chi N) < -1/8$) signals a smooth crossover between the two phases (see Methods). Curves for $\delta = 0$ (green solid line, V1),  $\Omega/(\chi N) = 0.2$ (blue solid line, H1), and for $\Omega/(\chi N) = 0.7$ (blue dashed line, H2) are shown on both diagrams and experimentally investigated in  Figs.~2(b), 3(a) and 3(b), respectively. The dependence of the transition point on both $\delta/(\chi N)$ and $\Omega/(\chi N)$ is investigated in Extended Data Fig.~2(b). \label{fig:PhaseDiagram}}\end{figure*}

Here we report an advance towards the goal of simulating quantum magnetism in an optical cavity. We observe a dynamical phase transition generated by coupling a narrow-linewidth optical transition of an ensemble of strontium atoms to a single detuned cavity mode [Fig.~\ref{fig:PhaseDiagram}a)i)].

In general terms, non-equilibrium phase transitions, characterized by the existence of a critical point that separates phases with distinct properties, have been described in various contexts. In driven open systems, non-equilibrium phases are signalled by different steady states that depend on system parameters such as pump or loss rates \cite{Marino,Barberena2019,Ritsch2017,Landin2018,Kroeze2018,Kroeze2019}, independent of initial conditions. Conversely, here we focus on a non-equilibrium phase transition in a closed system -- often referred to as a dynamical phase transition (DPT) -- where the non-equilibrium quantum phases are dynamical in nature: that is, qualitatively distinct behaviors are observed below, above or at a critical point \cite{Eckstein_2009,Schiro10,Sciolla,gambassi,Knap} in terms of the time average of an order parameter such as magnetization. DPTs are typically initiated by quenching control parameters and depend on the initial state of the system. Such DPTs have been observed experimentally in arrays of trapped ions \cite{Monroe_DPT_2017} and cold gases \cite{Thywissen_DPT_2018}, as well as previously in the context of macroscopic self-trapping \cite{Smerzi_1997,Oberthaler_Tunneling_2005,Abbarchi_2013,Reinhard_2013}. Here we demonstrate a DPT in a system of cold atoms with global interactions mediated by an optical cavity.

\section*{Implementation of the Lipkin-Meshkov-Glick Model}

A feature of our cavity simulator [Fig.~\ref{fig:PhaseDiagram}a)], compared with earlier observations, is the use of an ensemble of $N \approx 10^5$-$10^6$ cold atoms. We use two long-lived electronic levels in these atoms, $\ket{\downarrow}$ [$^1$S$_0$ ($m_J=0$)] and $\ket{\uparrow}$ [$^3$P$_1$ ($m_J=0$)] states, to mimic a spin-1/2 system ($\ket{\downarrow}$ and $\ket{\uparrow}$, respectively). The atoms are confined in a one-dimensional (1D) optical lattice with a near-magic-wavelength of $813$~nm supported by the optical cavity. We operate the experiment in a regime in which the atoms couple to a single common transverse electromagnetic (TEM$_{00}$) mode of the optical cavity with resonance frequency $\omega_c$ detuned by $\Delta=\omega_c-\omega_a $ from the atomic optical transition with frequency $\omega_a$. Here $\vert\Delta\vert$ is large with respect to the linewidths of the cavity, $\kappa/(2\pi) = 153.0(4)$~kHz, and atomic transition, $\gamma/2\pi = 7.5$~kHz, and also the vacuum Rabi splitting $g\sqrt{N}$ induced by the atoms, with $g$ the single-photon Rabi frequency $2g/(2\pi) = 21.8$~kHz. This means that the cavity-mediated dynamics of the atoms essentially conserves energy and can be well described by the following Hamiltonian: 
\begin{equation}
    \hat{H} = \hbar\chi \hat{J}^+\hat{J}^- + \hbar\Omega \hat{J}_x  - \hbar\delta \hat{J}_z . \label{eqn:Hamiltonian}
\end{equation} 
This Hamiltonian can be recast as the well-known Lipkin-Meshkov-Glick (LMG) model \cite{LMG,Vidal_LMG_2007}, which has been studied in various contexts, including quantum magnetism. In Eq.~\ref{eqn:Hamiltonian}, we have introduced the collective spin operators $\hat{J}_{\alpha} = \sum_j \hat{\sigma}^{\alpha}_j/2$, where $\hat{\sigma}^{\alpha}_j$ is a Pauli operator for the $j$th atom with $\alpha = x,y,z$ and $\hat{J}^{\pm} = \hat{J}_x \pm i\hat{J}_y$. The summation runs over the individual atoms $j=1,...,N$ in the cavity. The parameter $\chi$ sets the strength of the infinite-range exchange interactions mediated by the cavity mode, and $\Omega$ and $\delta$ define the strength of the transverse and longitudinal fields respectively (Fig.~\ref{fig:PhaseDiagram}a)]. The model is realized in the limit in which the cavity field couples identically to all atoms trapped in the optical lattice (see Methods for modifications due to inhomogeneity in this coupling).

\begin{figure*}[!htb]
\includegraphics[width = 6.75in]{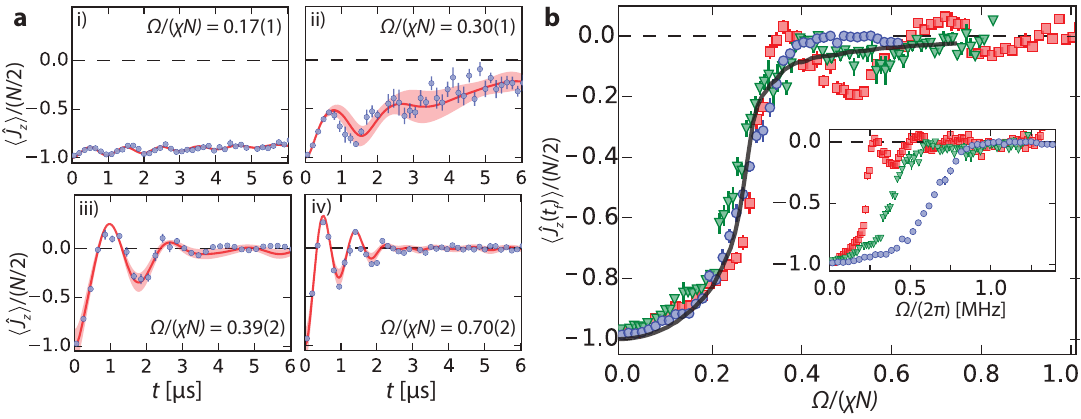}
\caption{{\bf Characteristic evolution of dynamical phases and scaling of DPT with atom number.} (a) Time-traces of the mean magnetization $\langle \hat{J}_z \rangle$ for the case of all spins initially in $\ket{\downarrow}$, $\delta = 0$ and $\Omega$ quenched to different values, at $t=0$, in the ferromagnetic [(i) and (ii)] and paramagnetic [(iii) and (iv)] phases for $N = 950\times10^3$ atoms and $\Delta/(2\pi) = 50~$MHz. The experimental data (blue) are compared to theoretical calculations (red lines) based on a mean-field description including relevant experimental details (see Methods and Supplementary Information). Shaded theoretical region accounts for shot-to-shot fluctuations in $\Omega/(\chi N)$. Each point is the average of 12 experimental repetitions. (b) Magnetization $\langle \hat{J}_z(t_f) \rangle$ for different numbers of atoms $N = (935,\ 620,\ 320)\times 10^3$ (blue, green and red, respectively) after $t_f = 4~\mu$s of evolution for different normalized drive strengths $\Omega/(\chi N)$ for $\Delta/(2\pi) = 50~$MHz and $\delta = 0$. This measurement maps to the green solid line (V1) in Fig.~\ref{fig:PhaseDiagram}(b) and (c). The drive strength normalization in each experimental shot is done by spin-dependent imaging. The solid black line indicates the simulated average (0 to 6~$\mu$s) as a function of the normalized drive including dephasing sources. The inset shows the magnetization versus non-normalized transverse field strength $\Omega$ for the same data sets. All error bars in experimental data are statistical ($1\sigma$).}
\label{fig:Nscale}
\end{figure*}

\section*{Dynamical Phase Diagram of the LMG model}

On varying the ratios between $\Omega$, $\delta$ and $\chi$, two distinct dynamical phases emerge, for which the time-averaged collective magnetization (along $\hat{z}$) of the atomic ensemble $\overline{\langle \hat{J}_z \rangle} \equiv \lim_{T\to\infty} (1/T)\int_0^T \langle \hat{J}_z(t)\rangle dt$ serves as an order parameter. When all spins are initially prepared in the $\ket{\downarrow}$ state and $\delta=0$, the system features a sharp second-order transition \cite{PhysRevB.99.045128} between a dynamical  ferromagnetic phase with $\overline{\langle \hat{J}_z \rangle} \neq 0$ and a dynamical paramagnetic phase with $\overline{\langle \hat{J}_z \rangle} = 0$. This transition is indicated by the solid green line (V1) on the phase diagram shown in Fig.~\ref{fig:PhaseDiagram}b), as well as its projection on the $\overline{\langle \hat{J}_z \rangle} $ vs $\Omega$ plane in the same panel and in Fig.~\ref{fig:PhaseDiagram}c). More generally, as a function of the parameters $\Omega$ and $\delta/(\chi N) \geq -1/8$, we observe a non-analyticity of the order parameter $\overline{\langle \hat{J}_z \rangle}$ [indicated by a solid white line in Fig.~\ref{fig:PhaseDiagram}c)], which marks a second-order transition between the two dynamical phases. However, the transition line is interrupted at a critical point [$\delta/(\chi N) = -1/8$]. Beyond this, there is a smooth crossover regime [indicated by a white dashed line in Fig.~\ref{fig:PhaseDiagram}c)] in which the system is ruled mainly by single-particle physics (set by  $\delta$ and $\Omega$) and has an intermediate behavior between that of a ferromagnet and a paramagnet.

In the ferromagnetic phase [red region in Fig.~\ref{fig:PhaseDiagram}b) and c)], the instantaneous magnetization $\langle \hat{J}_z \rangle$ oscillates about a non-zero time-averaged value, and the collective pseudospin Bloch vector $\langle\boldsymbol{\hat{J}}\rangle \equiv (\langle \hat{J}_x\rangle,\langle \hat{J}_y\rangle,\langle \hat{J}_z\rangle)$ remains trapped below the equator of the Bloch sphere throughout the dynamics. This phase is dominated by the interactions which can be understood in a mean-field approximation as $\chi\hat{J}^+\hat{J}^- \approx \chi ({\boldsymbol{\hat J}}\cdot {\boldsymbol{ \hat J}} -\hat{J}_z^2 )\approx \chi (N/2)^2 -2\chi\langle\hat{J}_z\rangle\hat{J}_z$. The term ${\boldsymbol{\hat J}}\cdot {\boldsymbol{ \hat J}}$ is a constant when restricted to the fully symmetric spin manifold, which is the case of interest here. The second term describes a self-induced precession of the collective Bloch vector about the $\hat{z}$-axis, which effectively tilts the axis of rotation of the comparatively weak transverse field, such that the trajectory of the Bloch vector deforms into an orbit that remains below the equatorial plane. 

Conversely, the paramagnetic phase [blue region in Fig.~\ref{fig:PhaseDiagram}b) and c)] is dominated by Rabi flopping driven by the transverse field $\Omega \hat{J}_x$. This term causes large oscillations of the instantaneous $\langle \hat{J}_z \rangle$, and, for $\delta= 0$, the collective Bloch vector breaks through the equatorial plane and rotates about the entire Bloch sphere. 

\begin{figure}[b]
\includegraphics{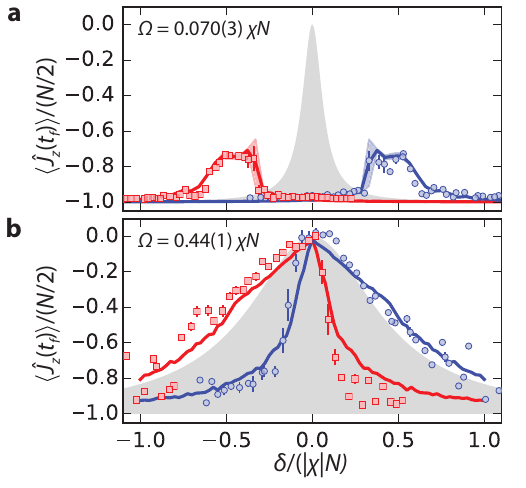}
\caption{{\bf Characterization of the DPT as a function of longitudinal field for two different transverse field values at fixed $\chi N$.} The atomic magnetization $\langle \hat{J}_z(t_f) \rangle$ at $t_f = 4~\mu$s is measured as a function of the normalized drive detuning $\delta/(\left|\chi\right| N)$ for cavity detunings $\Delta/(2\pi) = \pm 50~$MHz (red, $+50$~MHz; blue, $-50$~MHz) for two different drive strengths: (a) $\Omega = 0.070(3) \chi N$ and (b) $\Omega = 0.44(1) \chi N$. The inner edges of the resonant features in panel (a) indicate a sharp transition from ferromagnetic to paramagnetic phases as $\vert \delta \vert$ is increased. In contrast, the corresponding crossover in panel (b) is smoothed. Numerical simulations are shown as blue and red solid lines with corresponding shaded regions. The grey-shaded area indicates the non-interacting limit of Rabi-flopping. Measurements in (a) and (b) map, respectively, to cuts represented by the blue lines H1 and H2 in Fig.~\ref{fig:PhaseDiagram}(b)-(c). All error bars in experimental data  are statistical ($1\sigma$).}
\label{fig:detuning}
\end{figure}

For $\delta = 0$, the transition between the paramagnetic and ferromagnetic phases occurs at a critical drive $\Omega_c = \chi N/2$, as shown in Fig.~\ref{fig:PhaseDiagram}b) and c). The sharp transition in the dynamical behaviour of the system is traced back to the change in direction of the self-generated precession proportional to $\chi\langle\hat{J}_z\rangle\hat{J}_z$ as the Bloch vector crosses the equatorial plane at $\langle \hat{J}_z \rangle = 0$, generating an abrupt shift to large-amplitude oscillations for $\Omega > \Omega_c$. Typical dynamics of the collective Bloch vector in the ferromagnetic and paramagnetic phases are shown as insets in Fig.~\ref{fig:PhaseDiagram}b). The solid green (V1), solid blue (H1) and dashed blue (H2) lines in Fig.~\ref{fig:PhaseDiagram}b) and c) indicate analogous trajectories in the phase diagram which will be explored experimentally later in Figures \ref{fig:Nscale}b),~\ref{fig:detuning}a), and \ref{fig:detuning}b), respectively (see also Extended Data Fig.~2b and Supplementary Fig.~S3 for investigation of the transition as a function of detuning and drive).

\section*{Probing the LMG Dynamical Phase Diagram}

In our simulator, the cavity mediates a global spin-exchange interaction, which is microscopically described by a flip-flop process in which the emission of a photon from atom $i$ in state $\ket{\uparrow}$ into the cavity mode is subsequently absorbed by atom $j$ in state $\ket{\downarrow}$ [Fig.~\ref{fig:PhaseDiagram}a)iii)]. We operate in the regime $\vert\Delta\vert \gg g\sqrt{N}$, where the instantaneous average number of photons in the cavity mediating the interaction is much less than $N$, and the dynamics are well described by a spin-exchange model $\chi \hat{J}^+\hat{J}^-$ with coupling constant $\chi = -g^2/\Delta$ (see also Extended Data Fig.~2a and Methods). Similarly, the large detuning means that superradiant emission does not play an active role, in contrast to previous work \cite{Norcia_OAT_2018}. 
The interaction dynamics are faster than spontaneous emission, $\vert\chi\vert N \gg \gamma$, and satisfy the hierarchy $\vert\Delta\vert \gg g\sqrt{N} \gg \kappa,\gamma$. 

We realize the transverse fields $\Omega$ and $\delta$ by injecting laser light at frequency $\omega_p$ into the optical cavity through one mirror, creating a coherent driving field $\Omega_p e^{i\omega_p t}$ inside the cavity. In the rotating frame at $\omega_p$, the laser light's detuning from atomic resonance $\delta = \omega_p-\omega_a$ provides the longitudinal field $\delta \hat{J}_z$ in Eq.~\ref{eqn:Hamiltonian}. Moreover, the applied laser rapidly builds up a classical field within the cavity on a timescale of approximately $1/\Delta$, which couples $\ket{\downarrow}$ to $\ket{\uparrow}$. This realizes the transverse field $\Omega \hat{J}_x$ in Eq.~\ref{eqn:Hamiltonian}, where $\Omega = -2g\Omega_p/\Delta$. 
We adopt the convention that this transverse field is oriented along $\hat{x}$ in the pseudospin coordinate system such that by jumping the phase of the laser light, we are able to create transverse fields oriented along any direction in the pseudospin $x - y$ plane. Furthermore, the experiment is realized with a standing wave cavity, where incommensurate lattice and drive wavelengths generate inhomogeneous $\Omega$ and $\chi$ parameters compared with the ideal case presented above. This leaves unchanged the generic features of the phase diagram in Fig~\ref{fig:PhaseDiagram}, but quantitatively modifies the phase boundary.

\subsection*{DPT in the absence of a longitudinal field}

In Fig.~\ref{fig:Nscale}, we show experimental observations of the characteristic dynamics and DPT. We begin with all atoms in $\ket{\downarrow}$ and then quench $\Omega$ from zero to a specific value at $t=0$. After a variable evolution time, we rapidly freeze the atomic dynamics by quenching $\Omega\to 0$ and creating strong single-particle dephasing of the ground state. The atomic magnetization $\langle \hat{J}_z \rangle$ and atom number $N$ are then measured with high efficiency using fluorescence in combination with electron shelving and state-dependent displacements (see Methods and Extended Data Fig.~1).

For the time traces presented in Fig.~\ref{fig:Nscale}a), we map the magnetization across different drive strengths with fixed $\delta = 0$. For drives deep in the ferromagnetic phase [Fig.~\ref{fig:Nscale}(a)(i)], we observe small-amplitude oscillations that are in excellent agreement with our theoretical model based on a mean-field description of the system (see Methods and Supplementary Information). Close to the experimental critical point [Fig.~\ref{fig:Nscale}a)ii)-iii)], the dynamics become more complicated due to the complex interplay between interactions, drive and single-particle decoherence due to undesirable atomic motion in the optical lattice (see Supplementary Information and Supplementary Figs.~1 and 2). Deep in the paramagnetic phase [Fig.~\ref{fig:Nscale}a)iv)], we observe dynamics of the magnetization consistent with single-particle Rabi flopping with frequency $\Omega$ and in good agreement with our simulation. Damping of the oscillations occurs predominantly because of inhomogeneity in the coupling of the spins to the common cavity mode, shot-to-shot fluctuations in $\Omega/(\chi N)$ [attributed mostly to atom number fluctuations at about the $5\%$ (root mean square, r.m.s.) level] and atomic motion in the lattice. Spontaneous emission and decoherence related to leakage of photons from the cavity are negligible. We include these effects in our theoretical model [Fig.~\ref{fig:Nscale}a), red solid line], and fluctuations in $\Omega/(\chi N)$ are indicated by the red shaded regions. Typically, we notice that the experimentally calibrated parameters overestimate the value of $\Omega/(\chi N)$ by about 10\% compared with the numerical simulations. We attribute this systematic disagreement to drifts on the calibration parameters and details not captured by the theory model (see Supplementary Information).

We characterize the behaviour of the DPT with system size by measuring $\langle \hat{J}_z(t_f) \rangle$ at time $t_f = 4~\mu$s for different atom number $N$ while initializing every atom in $\ket{\downarrow}$ for $\delta = 0$ in Fig.~\ref{fig:Nscale}b). Measuring $\langle \hat{J}_z \rangle$ at a fixed time serves as a proxy of the long-term time-averaged magnetization, as considerable damping is caused by the previously mentioned effects. In the Fig.~\ref{fig:Nscale}b) inset, we observe a transition in the magnetization at different values of the transverse field $\Omega$, depending on atom number $N$. The dependence of the transition as a function of system size is demonstrated by re-scaling the corresponding drive as $\Omega/(\chi N)$, as shown in the main panel of Fig.~\ref{fig:Nscale}b), analogous to the green curve in Fig.~\ref{fig:PhaseDiagram}(b) and (c). We observe collapse of the data and a critical drive $\Omega^{\textrm{exp}}_c = 0.35(3) \chi N$. A comparison to theoretical calculations using time-averaged magnetization (see Methods) shows reasonable agreement (solid black line). The shift of the critical point relative to the ideal collective model, $\Omega_c/(\chi N)=1/2$, is predominantly attributable to the spatial inhomogeneity in the coupling of the atoms to the cavity mode (see Methods). Other small factors include single-particle decoherence of the atoms, which also contributes to the smearing out of the sharpness of the transition observed in the ideal system [Fig.~\ref{fig:PhaseDiagram}b)]. Nevertheless, a clear transition can be observed, as shown by comparing to the theoretical calculation.

\subsection*{DPTs at Fixed Transverse Fields}

The DPT can also be probed using our ability to controllably introduce a longitudinal field proportional to $\delta\hat{J}_z$ by detuning the injected light from the atomic transition, as shown in Fig.~\ref{fig:PhaseDiagram}b). In Fig.~\ref{fig:detuning}, we map out the response of the system to the drive detuning $\delta$ by measuring the order parameter $\langle \hat{J}_z(t_f) \rangle$ at $t_f=4~\mu$s for two fixed values of the drive strength $\Omega$ above and below the ($\delta = 0$) critical point, $\Omega^\mathrm{exp}_c$, and for two opposite cavity detunings $\Delta/(2\pi) = \pm 50$~MHz. 

We observe a sharp transition in the order parameter $\langle \hat{J}_z \rangle$ versus drive detuning, separating the ferromagnetic and the paramagnetic dynamical phases for a drive below the observed critical point $\Omega^{\mathrm{exp}}_c$ [blue solid line, H1, in Fig.~\ref{fig:PhaseDiagram}b) and c)]. This is plotted in Fig.~\ref{fig:detuning}a) with $\Omega = 0.070(3)\chi N < \Omega^{\mathrm{exp}}_c$. We observe sharp transitions at the inside edges of the resonant features, which occur symmetrically for each $\Delta$ at $\delta_c/\vert\chi N\vert = \mp 0.27(2)$. The critical value of $\delta_c$ and the gradual decrease in $\langle \hat{J}_z \rangle$ for large detuning show good agreement with a mean-field calculation. The robustness of the sharp transition is demonstrated by the symmetric response of the magnetization for $\Delta \leftrightarrow -\Delta$ and thus of the interaction shift $\chi N \leftrightarrow -\chi N$. 

\begin{figure}
\includegraphics[width=\columnwidth]{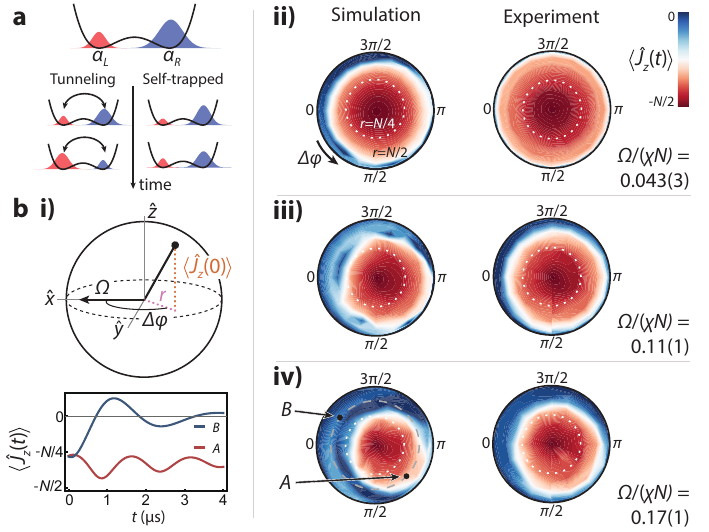}
\caption{{\bf Dependence of dynamical phases on initial conditions.} (a) The initial state on the Bloch sphere and subsequent dynamics of the spin model can be mapped to that of atomic condensates in a double-well potential, described by coherent complex amplitudes in the left and right wells, $\alpha_L$ and $\alpha_R$, respectively. Population imbalance maps to the magnetization ($J_z \sim |\alpha_L|^2-|\alpha_R|^2$), and the relative phase of the condensate wavefunctions maps to the azimuthal angle of the spin state ($\Delta\phi$). As time evolves, the population imbalance either oscillates as atoms tunnel back and forth between the wells (tunneling phase) or remains approximately constant (self-trapped phase). (b) (i) The initially prepared spin state can be parameterized in terms of the projection onto the equatorial plane $r = \sqrt{(N/2)^2 - \langle \hat{J}_z(0)\rangle^2}$ and the relative azimuthal phase $\Delta\phi$ between the initial collective Bloch vector and the coherent drive (see Methods). (ii)-(iv) Colour map of $\langle J_z(t_f) \rangle$ at $t_f=4~\mu$s of evolution, plotted in a polar projection of the Bloch sphere with coordinates defined by the initial condition as in panel (i) (initial conditions are always below the equator; they are shown above the equator in the figure to simplify visualization). Left (right) panels show simulated (experimental) results for $\langle J_z(t_f) \rangle$ at $t_f=4~\mu$s of evolution for different normalized drives $\Omega/(\chi N)$. The adjacent plot of $\langle \hat{J}_z(t) \rangle$ indicates typical dynamics in the red and blue regions.}
\label{fig:Icondition}
\end{figure}

Conversely, when the drive is tuned above $\Omega^\mathrm{exp}_c$, $\Omega = 0.44(1)\chi N > \Omega^{\mathrm{exp}}_c$ [Fig.~\ref{fig:detuning}b)], indicated by the blue dashed line, H2, in Fig.~\ref{fig:PhaseDiagram}b) and c), we observe a smoother crossover between the paramagnetic and ferromagnetic phases about the detuning $\delta_c/\vert\chi N\vert = \pm 0.04(3)$ in agreement with the mean-field calculation. Tuning $\delta < \delta_c$ ($\Delta > 0$) reduces the influence of the collective interactions and the magnetization resembles the prediction of single-particle detuned Rabi flopping. 

In both cases, the response of the system to $\delta$ can be understood by interpreting the single-particle shift and interaction in Eq.~(\ref{eqn:Hamiltonian}) as a nonlinear detuning proportional to $(2\chi\langle\hat{J}_z\rangle + \delta)\hat{J}_z$, which competes with the coherent drive. Depending on the sign of the interaction and the instantaneous magnetization, the single-particle term $\delta$ can either cancel or enhance the contribution of the interactions relative to the coherent drive, tuning the system between the ferromagnetic and paramagnetic dynamical phases. The predominant role of the interactions in the dynamics, especially below the critical point, can be observed by contrasting with the purely single-particle model of detuned Rabi oscillations (grey shaded area), which predicts a Lorentzian lineshape centered at $\delta = 0$. 

\section*{Sensitivity to initial condition}
The single-particle control achievable in our experimental platform allows us to explore the DPT as a function of the initial state, as shown in Fig.~\ref{fig:Icondition}. Specifically, we are able to demonstrate that the critical point of the transition is state-dependent, by preparing the collective pseudospin in different positions on the Bloch sphere. 
For example, we can prepare the system with $\Omega < \Omega_c^{\rm exp}$ such that the initial collective states near the south pole remained trapped below the equator, yet there also exists initial states prepared further towards the equator which exhibit large oscillations around the Bloch sphere characteristic of the paramagnetic phase.

Probing the response of the dynamics to different initial conditions allows us to establish a connection between the DPT in our effective spin model and the phenomena of macroscopic self-trapping and Josephson tunneling observed in coupled atomic condensates \cite{Oberthaler_Tunneling_2005} and solid state polariton condensates \cite{Abbarchi_2013}. Fig.~\ref{fig:Icondition}a) schematically shows a double-well atomic condensate, where the initial magnetization of the collective state on the Bloch sphere is analogous to the initial population imbalance between the wells, while the azimuthal angle maps to the relative phase difference of the condensates. Similarly, the ferromagnetic and paramagnetic phases can be related to the self-trapped and tunneling phases respectively \cite{Smerzi_1997}. 

In the rightmost panels [Fig.~\ref{fig:Icondition}b)ii)-iv)], we plot the measured magnetization after $4~\mu$s of evolution using a polar projection of the Bloch sphere for different drive strengths $\Omega$, as we scan the initially prepared state $\boldsymbol{J}(0)$. Here, the radial coordinate maps to the magnitude of the projection of $\boldsymbol{J}(0)$ on the equatorial plane (for $\langle \hat{J}_z(0) \rangle <0$), and the angle $\Delta\phi$ maps to the relative phase between the coherent drive and $\boldsymbol{J}(0)$ (see Methods). As we increase the drive strength, the set of initial conditions that lead to the ferromagnetic phase shrinks (red region) while also becoming increasingly asymmetric about the south pole. Both of these features are in qualitative agreement with our theoretical calculations, also shown in Fig.~\ref{fig:Icondition}(c)(ii)-(iv), which take into account coupling inhomogeneities, dephasing and shot-to-shot fluctuations on $\Omega/(\chi N)$. Quantitative differences arise predominantly due neglecting axial motion of the atoms in the theoretical model.

\section*{Conclusion}
The demonstration of the cooperation and competition between coherent drive and infinite-range interactions in an optical transition opens a path to the quantum simulation of richer spin models and out-of-equilibrium physics. For example, more complex spin-spin couplings can be engineered by using the available Zeeman sublevels of the $^3$P$_1$ state with two different cavity polarizations \cite{Davis_2019}.
Moreover, in the presence of additional inhomogeneous terms, our system can explore dynamical phases predicted to exist in Bardeen-Cooper-Schrieffer superconductors \cite{Barank04,Gurarie2015}, and by modulation of the transverse field our platform should be able to realize the archetypal model of a kicked top \cite{Swingle_2016}, relevant for explorations of quantum chaos and scrambling dynamics \cite{Swingle2018}. Lastly, our investigation of non-equilibrium dynamics using the $^{88}$Sr ($^1$S$_0$-$^3$P$_1$) optical transition can lead to insight into how to generate entangled states for quantum sensing with the long-lived $^{87}$Sr ($^1$S$_0$-$^3$P$_0$) optical transition used in state-of-the-art atomic clocks \cite{Campbell_2017}.

\bibliography{library}

\clearpage
\newpage

\setcounter{figure}{0}
\renewcommand{\thefigure}{ED\arabic{figure}}

\section*{Methods}

\subsection{Experimental description}

\begin{figure*}[t]
\includegraphics[width = 6.75in]{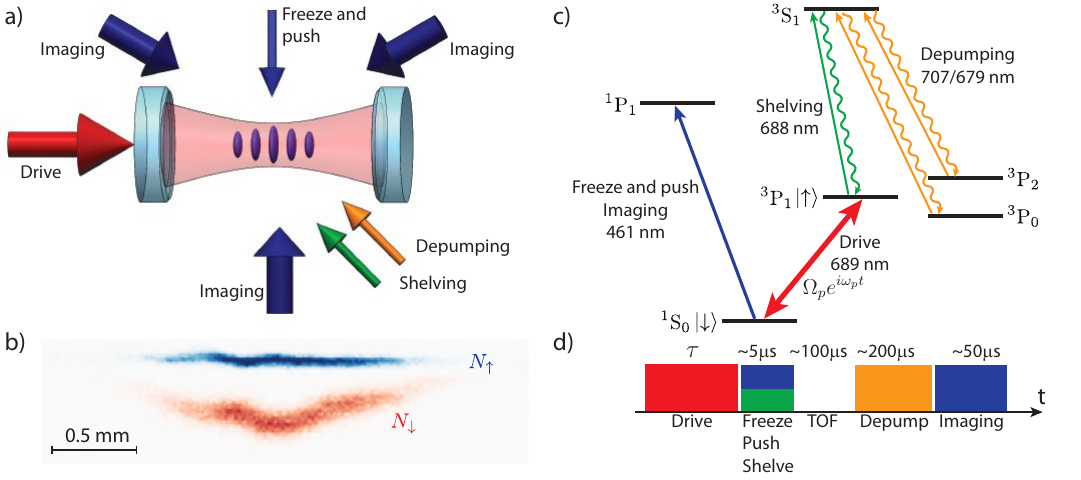}
\caption{{\bf Experimental platform.} (a) An optical cavity is driven by a $689$~nm coherent field  that establishes an intra-cavity field $\Omega_p e^{i\omega_p t}$, which is near resonance with the $^1$S$_0$ to $^3$P$_1$ transition in $^{88}$Sr. Inside the cavity, an ensemble of atoms is confined in a one-dimensional optical lattice at $813$~nm. Different lasers are applied for shelving excited state atoms into long-lived metastable excited states, for freezing the system dynamics, for applying a radiation pressure force that pushes ground states in a direction transverse to the cavity axis, for optically pumping atoms from long lived metastable excited states back to the ground state, and for fluorescence imaging of atoms in the ground state. (b) A typical fluorescence image captured on a CCD showing the state-resolved imaging technique. One sees that the $N_e$ excited state atoms that were shelved into $^3$P$_{0,2}$ while the freeze/push beam was applied remain near the trapping region. The $N_g$ ground state atoms are pushed away from the trapping region. Based on their spatial location, the atoms assigned to be in the excited (ground) state are shown in false color blue (orange). (c) We show the relevant energy levels for $^{88}$Sr, the laser wavelengths, and their functions. (d) Experimental timing sequence and typical time scales are shown.}
\label{fig:ED1}
\end{figure*}

Our experiment begins by loading up to $1\times10^6$ $^{88}$Sr atoms from a magneto-optical trap into a near magic-wavelength 813~nm one dimensional intracavity optical lattice, as we have shown in \cite{Norcia_OAT_2018,Norcia_ND_2016,Norcia_SR_2018,Norcia_SR_2016}. This lattice is nominally near magic with respect to the ultra-narrow millihertz $^1$S$_0\rightarrow^3$P$_0$ clock transition at 698~nm, but can be made near magic-wavelength for our optical transition at 689~nm ($m_J=0$ states) between $^1$S$_0\rightarrow^3$P$_1$ by setting the angle between the linear polarization of the lattice and the quantization axis, in order to reduce potential dephasing due to the transverse spreading of the atoms in a non-magic trap. We estimate residual inhomogeneous broadening due to the lattice to be below 2~kHz. The lattice spacing is incommensurate with the intracavity probe standing wave leading to inhomogeneous coupling to the cavity mode. A sketch of the system is shown in Fig.~\ref{fig:ED1}a). The atoms are laser cooled to $14~\mu$K and trapped in the optical lattice, with typical axial trap oscillation frequency $\omega_{\mathrm{trap}}/(2\pi) = 200~$kHz. The atom number is measured using florescence imaging on the dipole-allowed $^1$S$_0\rightarrow^1$P$_1$ transition at 461~nm, and it is calibrated by comparing it with the vacuum Rabi splitting when the cavity is on resonance with the atomic transition ($\Delta=0$), as detailed in Ref.~\cite{Norcia_ND_2016}. We determine $\Delta =0$ and $\delta=0$ from measurements of the symmetry of the collective vacuum Rabi splitting.  The measured cavity linewidth is $\kappa/(2\pi) = 153.0(4)$~kHz.  The cavity length is adjusted using PZTs, such that it can be kept at a detuning $\Delta$ during the experiment.

The cavity is driven for some time $\tau$ by a near resonant laser that realizes a coherent driving field $\Omega_p e^{i\omega_p t}$ in the cavity, as shown in Fig.~\ref{fig:ED1}, where $\Omega_p$ is related to the input power $P$ by the expression $\Omega_p=\sqrt{\kappa_m P/(2\hbar\omega_p)}$, with $\kappa_m = \kappa T_m/(T_m+T_L)$. Here, we define $T_m$ and $T_L$ as the single mirror power transmission and loss coefficients, 105~ppm and 23~ppm respectively. The drive is turned on and off in approximately 10~ns using an in-fiber electro-optical modulator (EOM), that creates a sideband at detuning $\delta$ while other frequency components are far from resonance and suppressed by being even further from resonance with the cavity mode. We apply a strong magnetic field perpendicular to the cavity axis to define the quantization axis. The probe light is polarized along the magnetic field direction such that the system is an effective two-level system  $\ket{\downarrow} = \ket{{}^1\mathrm{S}_0 , m_J = 0}$  and $\ket{\uparrow}=\ket{{}^3\mathrm{P}_1 ,m_J = 0}$ transition. For a more complete energy level diagram, see Fig.~\ref{fig:ED1}c).

The observation of the DPT requires that we be able to take a snapshot of the magnetization $\langle \hat{J}_z\rangle$ after some period of dynamical evolution. To achieve this, we have developed a technique to quickly freeze the dynamics and then apply state-dependent spatial displacements of the cloud such that the populations in the ground and excited states $N_{\downarrow}$ and $N_{\uparrow}$ are imaged onto two different regions of a CCD  [Fig.~\ref{fig:ED1}b)].

After the drive is applied for some time $\tau$, as shown in the time sequence in Fig.~\ref{fig:ED1}d), we turn off the coherent drive by extinguishing the applied EOM sideband. In order to effectively count atoms in both excited and ground state immediately after the drive, and freeze any dynamics that could be caused by spontaneous emission or the transient decay of the cavity field, we shine a strongly focused 461~nm beam along the $\hat{z}$ axis and apply a strong $688$~nm shelving beam. The 461~nm beam immediately stops the dynamics as it dephases the atoms, overwhelming the single particle rotation and any collective interactions. In addition, the 461~nm beam exerts a radiation pressure force that gives a momentum kick to the ground state atoms, causing them to move away from the trapping region.  Simultaneously, the shelving beam optically pumps excited state atoms to the metastable $^3$P$_{0,2}$ states, Fig.~\ref{fig:ED1}c). We apply the shelving pulse for 5~$\mu$s.  For scale, at $2\mu$s, we observe that $>90\%$ of the atoms in the excited state are shelved.

To finish our state dependent detection, we allow for a short time of flight ($\sim 100~\mu$s) so that the momentum kick applied to the ground state atoms is translated into a few 100~$\mu$m  displacement in space. We then optically pump the shelved atoms back to $^3$P$_{1}$ using 679~nm and 707~nm light applied for 200~$\mu$s. The atoms then decay to the ground state via single-atom decay with time constant 21~$\mu$s. We then perform fluorescence imaging for $50~\mu$s to observe the number of atoms in the two spatially resolved clouds as shown in Fig.~\ref{fig:ED1}b). This allows us to measure the magnetization $\langle \hat{J}_z \rangle = (N_{\uparrow}-N_{\downarrow})/(2(N_{\downarrow}+N_{\uparrow}))$ and total atom number $N = N_{\downarrow}+N_{\uparrow}$ in a single shot. We found that the whole process efficiency is above 98\%, limited mostly by the efficiency of the shelving process.

In Fig.~\ref{fig:detuning}, we change the drive detuning $\delta$ by changing the frequency of the rf applied to the EOM. The gray shaded area represents the rms amplitude for  Rabi oscillations without interactions, i.e. $\chi = 0$ in our model, and the corresponding rms magnetization is calculated simply as 
\begin{equation}
    \langle \hat{J}_z \rangle^{\mathrm{rms}}_{\chi=0}  = -\frac{1}{2}\frac{\Omega^2}{(\delta^2+\Omega^2)} -\frac{1}{2}.
\end{equation}  

In Fig.~\ref{fig:Icondition}, the initial state preparation is accomplished by preparing each spin in $\ket{\downarrow}$ and then rotating the spins with a strong drive $\Omega>\Omega_c$ for some chosen time. At this point, $t=0$, the system has acquired a magnetization $\langle \hat{J}_z(0) \rangle$. We then simultaneously shift the phase of the driving field by $\Delta\phi-\pi/2$ and its amplitude to some $\Omega < \Omega_c$ and evolve for a fixed time, typically $4~\mu$s.  The phase and amplitude jumps are accomplished by changing the phase and amplitude of the rf tone driving the EOM. We are then able to initialize the collective pseudospin Bloch vector at different positions on the Bloch sphere, such that $\langle \hat{J}_z(0) \rangle$ and $\Delta\phi$ define the polar and the azimuthal angles, respectively, as indicated in the figure in the main text. As the phase of the driving field naturally defines the $\hat{x}$ and $\hat{y}$ axes for the spin degree of freedom, our protocol can equivalently be viewed as preparing the collective Bloch vector at analogous positions on the pseudospin Bloch sphere.

\begin{figure*}[t]
\includegraphics[width = 6.75in]{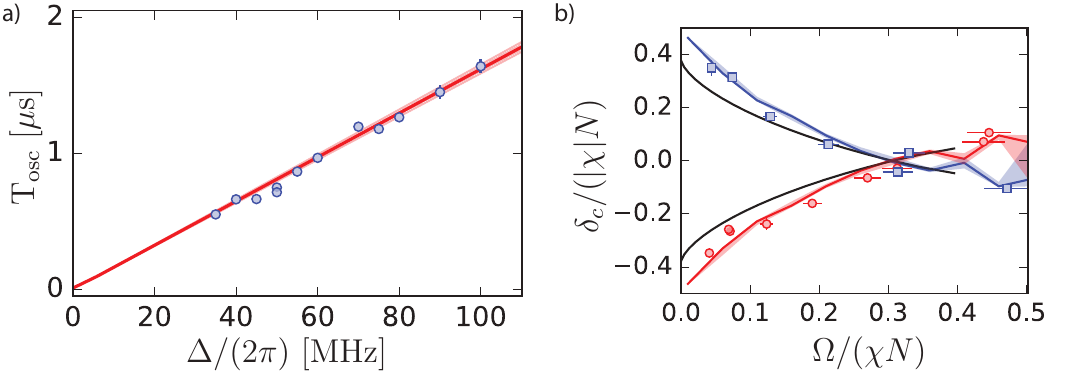}
\caption{{\bf Probing many-body dynamics and mapping the phase boundary.} (a) Oscillation period as function of the cavity detuning $\Delta$ for $2\Omega_p/(N g) = 0.104(4)$, $\delta = 0$ and atoms starting in $\ket{\downarrow}$. Blue markers are experimental values, solid red line represents the mean field prediction for the same drive and atom number, and the shaded red area represent typical experimental fluctuations on $2\Omega_p/(N g)$. The period is extracted from sinusoidal fits to data as in Fig.~\ref{fig:Nscale}(a), after removing a linear term caused by the single-particle dephasing effects. The mean field value, red solid line, is $\textrm{T}_\textrm{osc} = 2\pi/(N\chi)$ with the effective replacements due to inhomogeneous coupling as discussed in Methods. Measurements are taken in the dispersive limit where $\Delta\gg\sqrt{N}g$. (b) Critical detuning $\delta_c$ as function of the drive $\Delta$ for $\Delta/(2\pi) = \pm 50~$MHz (blue and red markers respectively). We also plot the theoretical prediction for the phase boundary [Eq.~(S30) of SI] with rescaled parameters, and predictions of the numerical model (solid lines) including uncertainty based on the typical fluctuations in $\Omega/(\chi N)$. Error bars are statistical (1$\sigma$).}
\label{fig:ED2}
\end{figure*}

\subsection{Model and simulations} \label{sec:methods_th}
The dynamics of the experimental system is modelled by a master equation for the density operator $\hat{\rho}$ of the complete atom-light system,
\begin{equation}
    \frac{d\hat{\rho}}{dt} = -\frac{i}{\hbar}\left[ \hat{H}_{\mathrm{tot}}, \hat{\rho} \right] + \mathcal{L}_c[\hat{\rho}] + \mathcal{L}_{el}[\hat{\rho}] + \mathcal{L}_s[\hat{\rho}]. \label{eqn:ALMasterEq_maintext}
\end{equation}
Here, the Hamiltonian $\hat{H}_{\mathrm{tot}} = \hat{H}_{\mathrm{A}} + \hat{H}_{\mathrm{L}} +  \hat{H}_{\mathrm{AL}}$ is split into three contributions characterizing the atoms, pumping of the cavity field and atom-light interaction respectively: 
\begin{eqnarray}
    \hat{H}_{\mathrm{A}} & = & \frac{\omega_a}{2} \sum_i \hat{\sigma}^z_i , \\
    \hat{H}_{\mathrm{L}} & = & \omega_c\hat{a}^{\dagger}\hat{a} + \Omega_p\left( \hat{a}e^{i\omega_p t} + \hat{a}^{\dagger}e^{-i\omega_p t} \right) , \\
    \hat{H}_{\mathrm{AL}} & = & \sum_i g_i \left( \hat{a}\hat{\sigma}^+_i + \hat{a}^{\dagger}\hat{\sigma}^-_i \right) ,
\end{eqnarray}
where $\hat{a}$ ($\hat{a}^{\dagger}$) is the annihilation (creation) operator of the cavity mode. To reiterate, $\omega_a$ is the frequency of the atomic transition, $\omega_c$ the frequency of the relevant cavity mode, $\Omega_p$ the effective amplitude of the injected field and $\omega_p$ the corresponding frequency. The spatial dependence of the coupling is characterized by $g_j = g\mathrm{cos}(k j)$ and $k = \pi \lambda_L/\lambda_c$, where $2g$ is the single-photon Rabi frequency at an anti-node of the cavity mode. This form arises because the magic wavelength of the 1D optical lattice $\lambda_L = 813$~nm is incommensurate with the wavelength $\lambda_c = 689$~nm of the cavity mode the atomic transition is coupled to. For simplicity, we take the summation to run over $i = 1,2,...,N$ total lattice sites, such that each site is assumed to be occupied by only a single atom. In reality, there are $\sim 10^3$ relevant lattice sites and each is occupied by $\sim 10^2$-$10^3$ atoms, however, as we assume contact interactions are not relevant and the atom-light coupling is consistent across the entire atomic sample this simplification is reasonable. 

Decoherence due to leakage of photons from the cavity at rate $\kappa$ is described by the Lindblad term
\begin{equation}
    \mathcal{L}_c[\hat{\rho}] = \frac{\kappa}{2} \left(2\hat{a}\hat{\rho}\hat{a}^{\dagger} - \hat{a}^{\dagger}\hat{a}\hat{\rho} - \hat{\rho}\hat{a}^{\dagger}\hat{a} \right) ,
\end{equation}
while spontaneous emission on the atomic transition at rate $\gamma$ and single-particle homogeneous broadening of the ensemble at rate $\gamma_{el}$ are described by
\begin{eqnarray}
    \mathcal{L}_s[\hat{\rho}] = \frac{\gamma}{2}\sum_i 2\hat{\sigma}^-_i\hat{\rho}\hat{\sigma}^+_i - \hat{\sigma}^+_i\hat{\sigma}^-_i\hat{\rho} - \hat{\rho}\hat{\sigma}^+_i\hat{\sigma}^-_i  , \\
    \mathcal{L}_{el}[\hat{\rho}] = \frac{\gamma_{el}}{2}\sum_i \hat{\sigma}^z_i \hat{\rho} \hat{\sigma}^z_i - \hat{\rho} .
\end{eqnarray}
The latter is attributed to a range of effects, including undesirable motion of the atoms in the optical lattice, and is discussed in more detail in the SI.

The simulations presented in Figs.~\ref{fig:Nscale}--\ref{fig:Icondition} are the result of numerical solution of Eq.~(\ref{eqn:ALMasterEq_maintext}) within the mean-field approximation [with the exception of the lower panels of Fig.~\ref{fig:Nscale}a) which include additional effects due to axial motion that are discussed in the SI]. Specifically, we solve equations of motion for $\boldsymbol{\sigma}_i \equiv (\langle \hat{\sigma}^x_i \rangle,\langle \hat{\sigma}^y_i\rangle,\langle \hat{\sigma}^z_i\rangle)$ and $\langle \hat{a} \rangle$, and factorize higher-order moments of the operators, e.g., $\langle \hat{\sigma}^x_i\hat{\sigma}^y_j \rangle \equiv \langle \hat{\sigma}^x_i \rangle \langle \hat{\sigma}^y_j \rangle$. Further details regarding the numerical simulations can be found in the SI.

The effective spin model which describes the nonlinear atomic dynamics throughout this manuscript is obtained from the atom-light model [Eq.~(\ref{eqn:ALMasterEq_maintext})] by separate adiabatic elimination of the injected field and intracavity fluctuations, and the full calculation is detailed in the SI. Here, we merely present the resulting Hamiltonian for the atoms:
\begin{equation}
    \hat{H} = \hbar\sum_{i,j} \chi_{ij} \hat{\sigma}^+_i \hat{\sigma}^-_j + \hbar\sum_i \frac{\Omega_i}{2} \hat{\sigma}^x_i - \frac{\hbar\delta}{2}\sum_i \sigma^z_i , \label{eqn:InhomHamiltonian_maintext}
\end{equation}
where $\chi_{ij} = -g_i g_j/\Delta$, $\Omega_i = -2g_i\Omega_p/\Delta$ with $\delta = \omega_p - \omega_a$ and $\Delta = \omega_c - \omega_a$. Moreover, we have assumed $\vert\Delta \vert \gg \kappa, g\sqrt{N}, \sqrt{g\Omega_p}, \delta$. In the limit $k = 2n\pi$ for $n\in\mathbb{Z}$, i.e. uniform atom-light coupling $g_j \to g$, then we recover the collective XY model of Eq.~\ref{eqn:Hamiltonian}.

Although in the experimental platform the atom-light coupling $g_j$ is spatially varying due to the incommensurate cavity and lattice wavelengths, the qualitative physics we explore is still consistent with the framework of the collective XY model. Specifically, while the simulations of Fig.~\ref{fig:Nscale}--\ref{fig:Icondition} take the proper form of $g_j$ into account (see SI), we observe that features of the detailed inhomogeneous model such as the critical point and dynamical time-scales are consistent with the collective model upon a rescaling of the atom-light coupling.

For weak drives deep in the paramagnetic phase the collective model replicates the quantitative predictions of the inhomogeneous model upon replacement of the atom-light coupling with the rms average, $g\to g/\sqrt{2}$ and thus $\chi \to \chi/2$ and $\Omega \to \Omega/\sqrt{2}$. This approximation is supported by comparison to experimental results for the period of the weak oscillations deep in the ferromagnetic phase, which are expected to be proportional to $1/(\chi N)$. In Extended Data Fig.~\ref{fig:ED2}a) we extract this period from the experimental data as a function of cavity detuning $\Delta$, which is equivalent to varying the interaction strength $\chi \propto 1/\Delta$. We confirm the fitted slope agrees with the $\chi \to \chi/2$ correction for inhomogeneous atom-light coupling.

As the drive is increased the rescaling required for quantitative comparison changes. Specifically, comparing to the critical point $\Omega^\textrm{theory}_c/(\chi N)$ obtained from a numerical calculation of the inhomogeneous model in the absence of decoherence we find that the corresponding collective model requires a rescaling $g\to 0.62g$, and thus $\chi \to 0.38\chi$ and $\Omega \to 0.62\Omega$, to match the critical value $\Omega^\textrm{theory}_c/(\chi N) \approx 0.31$. The reduction of this value below the true collective critical drive $\Omega_c/(\chi N) = 1/2$ is consistent with that observed experimentally ($\Omega^{\textrm{exp}}_c/(\chi N) = 0.35(3)$).

\subsection{Mapping the phase boundary}

In Fig.~\ref{fig:PhaseDiagram}b)-c) we present the system phase diagram (under the assumption of uniform atom-light coupling), where we map the magnetization $\langle \hat{J}_z \rangle$ as a function of the probe detuning $\delta$ and drive amplitude $\Omega$. A sharp boundary separates the dynamical phases for $\Omega/(\chi N) \lessapprox 0.65$, shown by the solid white line in Fig.~\ref{fig:PhaseDiagram}c). However, as the drive is further increased and for $\delta/(\chi N) <-1/8$ the boundary dilutes to a smooth crossover, as shown by the dashed white line in  Fig.~\ref{fig:PhaseDiagram}c). 

Using similar results as the ones shown in Fig.~\ref{fig:detuning}, for the inhomogeneous case relevant for experiment we are able to map out this boundary and define a critical detuning $\delta_c$ between the two dynamical phases for different fixed drive strengths $\Omega$. We identify these values by looking at the maximum gradient on each of the experimental and numerical $\langle \hat{J}_z \rangle$ against $\delta$ plots shown in Fig.~\ref{fig:detuning}. 
In Fig.~\ref{fig:ED2}b) we plot $\delta_c$ against $\Omega$ (markers) and compare to numerical simulations (solid lines) for two opposite cavity detunings $\Delta/(2\pi) =\pm50$~MHz. For values above $\Omega/(\chi N) \approx 0.31$ the solid traces do not represent a strict phase boundary but rather characterize the crossover region analogous to the crossover region indicated by a white  dashed line  in Fig.~\ref{fig:PhaseDiagram}c) for the  homogeneous case.

In the SI we derive an expression for the boundary between the two dynamical phases based on the model presented in the main text [Eq.~(\ref{eqn:Hamiltonian})] in the mean field limit. In the homogeneous case the phase boundary $\Omega_c(\delta)$ is, for $\delta/(\chi N)>-1/8$:
\begin{widetext}
\begin{equation}\label{eqn:MCriticalOmega}
    \frac{\Omega_c(\delta)}{\chi N}=\frac{1}{2}\Bigg[2\bigg(1-\frac{\delta}{\chi N}\bigg)\bigg(1+\frac{2\delta}{\chi N}\bigg)-\frac{3}{2}\bigg(\frac{8\delta}{\chi N}+1\bigg)+\frac{1}{2}\bigg(1+\frac{8\delta}{\chi N}\bigg)^{3/2}\Bigg]^{1/2}.
\end{equation}
\end{widetext}
To address the inhomogeneous coupling present in our experiment, we rescale $g\to 0.62g$ and thus $\chi \to 0.38\chi$ and $\Omega \to 0.62\Omega$ 
in this equation as described earlier. A comparison of the rescaled Eq.~(\ref{eqn:MCriticalOmega}) to the experimental data is shown as the black traces in Fig.~\ref{fig:ED2}b). 

\section*{Acknowledgements} We acknowledge helpful discussions with Ian Spielman, Murray Holland and Athreya Shankar. This work is supported by the AFOSR grant FA9550-18-1-0319 and its MURI Initiative, by the DARPA  Extreme Sensing and ARO grant W911NF-16-1-0576, the ARO single investigator award W911NF-19-1-0210,  the NSF PHY1820885, NSF JILA-PFC PHY-1734006 grants, and by NIST. J.~R.~K.~C. acknowledges financial support from NSF GRFP.

\section*{Author contributions} J.~A.~M., D.~J.~Y, J.~R.~K~.C. and J.~K.~T collected and analyzed the experimental data. R.~J.~L-S, D.~B. and A.~M.~R developed the theoretical model. All authors discussed the results and contributed to the preparation of the manuscript. 

\section*{Author information} The authors declare no competing interests.  Requests for reprints and materials should be made to the corresponding authors (A.~M.~R and J.~K.~T) at arey@jila.colorado.edu or jkt@jila.colorado.edu 

\section*{Data availability}
Data relevant to the figures and conclusions of this manuscript are available at https://doi.org/10.5061/dryad.mgqnk98w9 \cite{Dataset}

\clearpage
\newpage

\renewcommand{\theequation}{S\arabic{equation}}
\setcounter{equation}{0}
\renewcommand{\thesection}{S\arabic{section}}
\setcounter{section}{0}
\renewcommand{\thetable}{S\arabic{table}}
\setcounter{table}{0}
\renewcommand{\figurename}{\textbf{Fig.}}
\renewcommand{\thefigure}{\textbf{S\arabic{figure}}}
\setcounter{figure}{0}
\renewcommand{\figurename}{\textbf{Fig.}}

\onecolumngrid
\begin{center}
    \textbf{\Large{Supplementary Information}}
\end{center}

\begin{adjustwidth}{2 cm}{2 cm}
In this accompanying Supplementary Information we construct a detailed theoretical model describing the light-matter interaction in the cavity-QED platform. We elaborate on how the effective spin model arises from the underlying microscopic model, whilst also investigating the corrections to this model which arise due to, e.g., axial motion of the finite temperature atomic ensemble. Moreover, we present additional experimental characterization of these effects which display good agreement with numerical calculations.
\newline
\newline
\end{adjustwidth}

\twocolumngrid
\section{Effective spin model \label{sec:SpinModel}} 

The dynamics of the driven-dissipative atom-light system can be described by a Lindblad master equation for the density matrix $\hat{\rho}$, 
\begin{equation}
    \frac{d\hat{\rho}}{dt} = -\frac{i}{\hbar}\left[ \hat{H}_{\mathrm{tot}}, \hat{\rho} \right] + \mathcal{L}_c[\hat{\rho}] + \mathcal{L}_{el}[\hat{\rho}] + \mathcal{L}_s[\hat{\rho}]. \label{eqn:ALMasterEq}
\end{equation}
Here, the Hamiltonian $\hat{H}_{\mathrm{tot}} = \hat{H}_{\mathrm{A}} + \hat{H}_{\mathrm{L}} +  \hat{H}_{\mathrm{AL}}$ is split into three contributions characterizing the atoms, cavity field and atom-light interaction respectively: 
\begin{eqnarray}
    \hat{H}_{\mathrm{A}} & = & \hbar\omega_a\sum_i \hat{\sigma}^+_i\hat{\sigma}^-_i , \\
    \hat{H}_{\mathrm{L}} & = & \hbar\omega_c\hat{a}^{\dagger}\hat{a} + \hbar\left( \Omega_p^*\hat{a}e^{i\omega_p t} + \Omega_p\hat{a}^{\dagger}e^{-i\omega_p t} \right) , \label{eqn:HL} \\
    \hat{H}_{\mathrm{AL}} & = & \hbar\sum_i g_i \left( \hat{a}\hat{\sigma}^+_i + \hat{a}^{\dagger}\hat{\sigma}^-_i \right) , 
\end{eqnarray}
where $\hat{a}$ ($\hat{a}^{\dagger}$) is the annihilation (creation) operator of the cavity mode and $\hat{\sigma}^{\pm,z}_i$ are the Pauli operators at site $i$. We assume the cavity drive $\Omega_p = \vert \Omega_p \vert e^{i\phi}$ has a well defined amplitude $\vert \Omega_p \vert$ and phase $\phi$. 

Our model assumes atoms are  frozen in place at different  sites of a  1D optical lattice (see Methods for further discussion). The interaction between an atom at site $i$ and the light  is parametrized  by  the single-photon Rabi frequency $2g_i = 2g\mathrm{cos}(k i)$. The spatial variation of the coupling is characterized by $k = \pi\lambda_L/\lambda_c$, with $\lambda_L = 813$~nm the magic wavelength of the confining lattice and $\lambda_c = 689$~nm the cavity wavelength. Decoherence due to leakage of photons from the cavity at rate $\kappa$ is described by the Lindblad term
\begin{equation}
    \mathcal{L}_c[\hat{\rho}] = \frac{\kappa}{2} \left(2\hat{a}\hat{\rho}\hat{a}^{\dagger} - \hat{a}^{\dagger}\hat{a}\hat{\rho} - \hat{\rho}\hat{a}^{\dagger}\hat{a} \right) ,
\end{equation}
while spontaneous emission on the atomic transition and single-particle homogeneous broadening of the ensemble are described by
\begin{eqnarray}
    \mathcal{L}_s[\hat{\rho}] = \frac{\gamma}{2}\sum_i 2\hat{\sigma}^-_i\hat{\rho}\hat{\sigma}^+_i - \hat{\sigma}^+_i\hat{\sigma}^-_i\hat{\rho} - \hat{\rho}\hat{\sigma}^+_i\hat{\sigma}^-_i  , \\
    \mathcal{L}_{el}[\hat{\rho}] = \frac{\gamma_{el}}{2}\sum_i \hat{\sigma}^z_i \hat{\rho} \hat{\sigma}^z_i - \hat{\rho} .
\end{eqnarray}
The latter is attributed to a range of effects, including: motion of the atoms in the optical lattice and background collisions. Other effects such as phase noise on our drive laser and broadening due to a non-magic optical lattice have been addressed experimentally and reduced as much as possible.

The effective spin model for the atomic dynamics which is presented in the manuscript [Eqs.~(1) and (10)] is obtained from Eq.~(\ref{eqn:ALMasterEq}) by a sequence of frame transformations and separate adiabatic elimination of the injected field and intracavity fluctuations. Briefly, we first move to the rotating frame $\hat{a} \to \hat{a}e^{i\omega_p t}$ and then $\hat{\sigma}^+_i \to \hat{\sigma}^+_i e^{-i\omega_p t}$. The master equation becomes 
\begin{equation}
    \frac{d\hat{\rho}}{dt} = -\frac{i}{\hbar}\left[ \hat{H}_{\mathrm{eff}}, \hat{\rho} \right] + \mathcal{L}_c[\hat{\rho}] + \mathcal{L}_{el}[\hat{\rho}] + \mathcal{L}_s[\hat{\rho}]. \label{eqn:ALMasterEqnStep2}
\end{equation}
where the effective Hamiltonian is
\begin{multline}
    \hat{H}_{\mathrm{eff}} = \hbar(\Delta - \delta)\hat{a}^{\dagger}\hat{a} + \hbar(\Omega_p^*\hat{a} + \Omega_p\hat{a}^{\dagger}) \\
    -\frac{\hbar\delta}{2}\sum_i \hat{\sigma}^z_i + \hbar\sum_i  g_i \left( \hat{a}\hat{\sigma}^+_i + \hat{a}^{\dagger}\hat{\sigma}^-_i \right), 
\end{multline}
where we have defined the cavity detuning $\Delta=\omega_c-\omega_a$ and the atomic detuning $\delta=\omega_p-\omega_a$. The intracavity field can be decomposed into a classical term describing the injected light and residual quantum fluctuations: $\hat{a} = \alpha + \hat{a}^{\prime}$. To derive the final effective spin model, we will adiabatically eliminate the classical and quantum contributions of the intracavity field individually. The classical field $\alpha$ can be found by exact solution of the equation of motion $\dot{\alpha} = -i(\Delta - \delta - i\kappa/2)\alpha - i\Omega_p$, which is obtained by equating $\dot{\alpha}$ with terms $\propto \alpha$ and the drive $\Omega_p$ in the mean-field equation $\partial_t\langle\hat{a}\rangle$ obtained from the master equation [Eq.~(\ref{eqn:ALMasterEqnStep2})]. As we will focus on time-scales associated with the atomic degrees of freedom, using $\vert\Delta - \delta\vert \gg \sqrt{g\Omega_p}$ we average the fast oscillating contributions in the analytic solution and find $\alpha \approx -2\Omega_p/[2(\Delta - \delta) - i\kappa]$. This choice allows us to eliminate the coherent drive from the master equation, while we will account for the atomic back-action in the later elimination of $\hat{a}^{\prime}$. Specifically, substitution of $\hat{a} = \alpha + \hat{a}^{\prime}$ back into Eq.~(\ref{eqn:ALMasterEqnStep2}) with this solution yields
\begin{equation}
    \frac{d\hat{\rho}}{dt} = -\frac{i}{\hbar}\left[ \hat{H}^{\prime}_{\mathrm{eff}}, \hat{\rho} \right] + \mathcal{L}^{\prime}_c[\hat{\rho}] + \mathcal{L}_{el}[\hat{\rho}] + \mathcal{L}_s[\hat{\rho}] ,
\end{equation}
where 
\begin{multline}
    \hat{H}^{\prime}_{\mathrm{eff}} = \hbar(\Delta - \delta)\hat{a}^{\prime\dagger}\hat{a}^{\prime} - \frac{\hbar\delta}{2}\sum_i \hat{\sigma}^z_i \\
    + \hbar\sum_i \left[ \frac{\Omega_i}{2}\hat{\sigma}^x_i + \frac{\Omega^{\prime}_i}{2}\hat{\sigma}^y_i + g_i \left( \hat{a}^{\prime}\hat{\sigma}^+_i + \hat{a}^{\prime\dagger}\hat{\sigma}^-_i \right) \right].
\end{multline}
Here, we have defined the effective fields
\begin{eqnarray}
 \Omega_i & = & \frac{g_i\vert\Omega_p\vert \left[ \kappa\mathrm{sin}(\phi) - 2(\Delta-\delta)\mathrm{cos}(\phi) \right]}{(\Delta - \delta)^2 + (\kappa/2)^2}  , \\
 \Omega^{\prime}_i & = & \frac{g_i\vert\Omega_p\vert \left[ \kappa\mathrm{cos}(\phi) + 2(\Delta-\delta)\mathrm{sin}(\phi) \right]}{(\Delta - \delta)^2 + (\kappa/2)^2}  ,
\end{eqnarray}
and we remind the reader that $\phi$ is the phase of the cavity drive, $\Omega_p = \vert \Omega_p \vert e^{i\phi}$.From the above expressions that
when the empty cavity is  driven on resonance, the external  laser  establishes a coherent state inside the cavity with average intracavity photon number   $(2 \Omega_p/\kappa)^2$.

Now, the photon loss   with respect to the fluctuating component is  accounted for by the Lindblad term
\begin{equation}
    \mathcal{L}_c[\hat{\rho}] = \frac{\kappa}{2} \left(2\hat{a}^{\prime}\hat{\rho}\hat{a}^{\prime\dagger} - \hat{a}^{\prime\dagger}\hat{a}^{\prime}\hat{\rho} - \hat{\rho}\hat{a}^{\prime\dagger}\hat{a}^{\prime} \right) .
\end{equation}

When the cavity is far-detuned from the pump frequency, $\vert\Delta - \delta\vert = \vert\omega_c - \omega_p \vert \gg g\sqrt{N}$, the bosonic mode $\hat{a}^{\prime}$ may be adiabatically eliminated \cite{Norcia_OAT_2018,Haake_1971}. This results in an effective spin model, described by a master equation for the reduced density matrix of the atoms $\hat{\rho}_a$:
\begin{equation}
    \frac{d\hat{\rho}_a}{dt} = -\frac{i}{\hbar}\left[ \hat{H}, \hat{\rho}_a \right] + \mathcal{L}_{el}[\hat{\rho}_a] + \mathcal{L}_s[\hat{\rho}_a] ,
\end{equation}
with 
\begin{equation}
    \hat{H} = \hbar\sum_{i,j} \chi_{ij} \hat{\sigma}^+_i \hat{\sigma}^-_j + \hbar\sum_i \frac{\Omega_i}{2} \hat{\sigma}^x_i + \hbar\sum_i \frac{\Omega^{\prime}_i}{2} \hat{\sigma}^y_i - \frac{\hbar\delta}{2}\sum_i \sigma^z_i , \label{eqn:InhomHamiltonian}
\end{equation}
for $\chi_{ij} = -g_i g_j\Delta/[\Delta^2 + (\kappa/2)^2] \approx -g_i g_j/\Delta$, where the latter approximation is valid for $\kappa \ll \vert \Delta \vert$. We have ignored additional collective spontaneous emission which can arise in this treatment \cite{Norcia_OAT_2018} as we operate in the limit of a large cavity detuning $\Delta \gg \delta,\kappa$. Excluding the state preparation protocol used in Fig.~4 of the manuscript, we can set $\phi = 0$ and thus under the same limit of large detuning we have $\Omega_i \approx -2g_i\vert\Omega_p\vert/\Delta$ and $\Omega^{\prime}_i \approx 0$, corresponding to Eq.~(10) of the Methods.

From inspection of this derivation, it becomes clear that the external pumping of the cavity mode coherently drives the atomic transition via the term $\propto \hat{\sigma}^x_i$, while the quantum fluctuations of the cavity field mediate effective spin-spin interactions $\propto \hat{\sigma}^+_i\hat{\sigma}^-_j$ between atoms $i$ and $j$. In particular, the latter interactions can be interpreted as arising because of the back-action of the atoms on the quantum fluctuations of the cavity mode.

\section{Axial motion \label{sec:AxialMotion}}
The effective spin model derived in the previous section specifically assumes that the atoms are localized at anti-nodes of the optical lattice and their motional degrees of freedom are completely frozen.
However, this description can be insufficient when the atomic transition is strongly driven by the externally injected field at rates much larger than the characteristic axial trap frequency. In this section we derive a simple theoretical model which incorporates axial motion within the optical lattice, and demonstrate how this motion leads to quantitative corrections to the predictions of the spin model for dynamics in the normal phase.

The atom-light interaction incorporating axial motion is more generally described as
\begin{equation}
    \hat{H}_{\mathrm{AL}} =\hbar \int dx ~ \left\{ g(x)  \hat{a} \hat{\psi}^{\dagger}_{\uparrow}(x) \hat{\psi}_{\downarrow}(x) + h.c. \right\} \label{eqn:HAL_motiongeneral}
\end{equation}
Here, $\hat{\psi}_{\sigma}(x)$ [$\hat{\psi}^{\dagger}_{\sigma}(x)$] is the atomic field operator describing annihilation (creation) of a boson with pseudospin $\sigma=\uparrow,\downarrow$ at axial position $x$ along the cavity, $g(x) = g\mathrm{cos}(k_cx)$ is the spatially-dependent atom-light coupling and $k_c = 2\pi/\lambda_c$ is the wave-vector of the cavity mode.

Treating each site of the optical lattice within the harmonic approximation \cite{Jaksch_1998}, the atomic field operator can be expanded in the basis of harmonic oscillator wavefunctions $\varphi_n(x)$,  $\hat{\psi}_{\sigma}(x) \equiv \sum_{j,n} \hat{b}_{j,n,\sigma} \varphi_n(x-x_j)$, where $j$ indexes the lattice site, $n$ the harmonic oscillator level, and $\hat{b}_{j,n,\sigma}$ the associated bosonic annihilation operator. Substituting this expansion into Eq.~(\ref{eqn:HAL_motiongeneral}) and keeping only single-site atomic processes yields
\begin{equation}
    \hat{H}_{\mathrm{AL}} = \sum_{n,m,j} \hbar g^{nm}_{j} \left( \hat{a}\hat{b}^{\dagger}_{j,n,\uparrow}\hat{b}_{j,m,\downarrow} + h.c. \right) , \label{eqn:HAL_motion}
\end{equation}
where the mode-dependent atom-light coupling can be expressed as 
\begin{equation}
    g^{nm}_j = g\mathrm{cos}(kj)\eta^{nm}_c + g\mathrm{sin}(kj)\eta^{nm}_s ,
\end{equation}
where $k = \pi\lambda_L/\lambda_c$ as previously defined. The coefficients $\eta^{nm}_c = \int dx ~ \mathrm{cos}(k_c x) \varphi_n(x)\varphi_m(x)$ and $\eta^{nm}_s = \int dx ~ \mathrm{sin}(k_c x) \varphi_n(x)\varphi_m(x)$ depend on the axial extent of the atomic wavefunction compared to the wavelength of the cavity mode. 

In terms of the bosonic operators for the atoms at each lattice site, we can then express the full dynamics of the system by the atom-light master equation 
\begin{equation}
    \frac{d\hat{\rho}}{dt} = -\frac{i}{\hbar}\left[ \hat{H}_{\mathrm{mot}}, \hat{\rho} \right] + \mathcal{L}_c[\hat{\rho}] + \mathcal{L}_{el}[\hat{\rho}] + \mathcal{L}_s[\hat{\rho}]. \label{eqn:MotionMasterEq}
\end{equation}
where $\hat{H}_{\mathrm{mot}} = \hat{H}_A + \hat{H}_L + \hat{H}_{\mathrm{AL}}$ as previous, with $\hat{H}_{\mathrm{AL}}$ and $\hat{H}_{\mathrm{L}}$ as per Eqs.~(\ref{eqn:HAL_motion}) and (\ref{eqn:HL}) respectively and 
\begin{multline}
    \hat{H}_{\mathrm{A}} = \frac{\hbar \omega_a}{2}\sum_{n,j} \left( \hat{b}^{\dagger}_{n,j,\uparrow} \hat{b}_{n,j,\uparrow} - \hat{b}^{\dagger}_{n,j,\downarrow} \hat{b}_{n,j,\downarrow}  
    \right)  \\
    + \sum_{n,j,\sigma} \hbar \omega_  {\mathrm{trap}} n \hat{b}^{\dagger}_{n,j,\sigma} \hat{b}_{n,j,\sigma} . \label{eqn:HAmotion}
\end{multline}
Here, $\omega_{\mathrm{trap}} = \sqrt{4V_0E_r}/\hbar$ is the effective trap frequency in the harmonic approximation, with $V_0$ the optical lattice potential height and $E_r = \hbar^2k_r^2/(2m)$ the recoil energy with $k_r$ the recoil momentum.

The level-changing process included in our model ($g^{nm}_j$ with $n\neq m$) become important when the effective drive of the atoms $\Omega$ becomes comparable to the trapping frequency. In particular, this theoretical model including axial motion is used to explain the experimental data in Fig.~2a) of the main text. 

For completeness, our calculations incorporating axial motion involve numerical solution of Eq.~(\ref{eqn:MotionMasterEq}) within the mean-field approximation. Specifically, we compute equations of motion for the expectations $\langle \hat{b}^{\dagger}_{n,j,\sigma} \hat{b}_{n,j,\sigma^{\prime}} \rangle$ and assume all higher-order correlations factorize into these expectations, whilst the cavity field is assumed to be uncorrelated: i.e.  $\langle \hat{a} \hat{b}^{\dagger}_{n,j,\sigma} \hat{b}_{n,j,\sigma^{\prime}} \rangle  \approx \langle \hat{a} \rangle \langle \hat{b}^{\dagger}_{n,j,\sigma} \hat{b}_{n,j,\sigma^{\prime}} \rangle$ and similar for other combinations. The cavity field is modelled with the expansion $\langle \hat{a} \rangle = \alpha + \langle \hat{a}^{\prime} \rangle$ as in the spin-only model of Sec.~\ref{sec:SpinModel}, where $\alpha = -2\Omega_p/[2(\Delta - \delta) - i\kappa]$. The fluctuating component is assumed to be adiabatically eliminated by setting $d\langle\hat{a}^{\prime}\rangle/dt = 0$. Solving this equation we find $\langle \hat{a}^{\prime} \rangle = -2\sum_{n,m,j} g^{n,m}_j \langle \hat{b}^{\dagger}_{j,m,\downarrow} \hat{b}_{j,n,\uparrow} \rangle/(2(\Delta-\delta) - i\kappa)$ and substitute this into the remaining equations of motion for the atomic moments, which are subsequently solved numerically.

\begin{figure*}[htb!]
\includegraphics[width = 6.75in ]{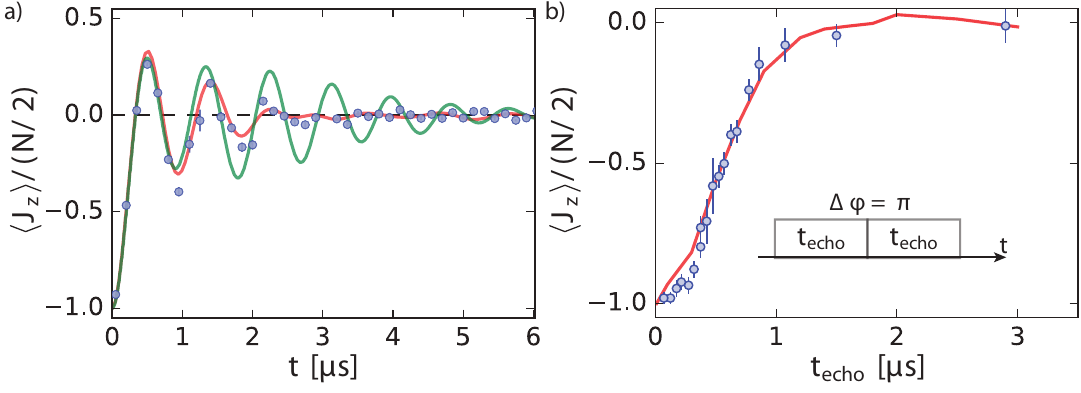}
\caption{{\bf Effects of axial motion on cavity-mediated dynamics.} (a) Comparison of theoretical models. Blue markers are observed experimental data for $\Omega/(\chi N) = 0.70(2)$ against numerical simulations from the mean-field dynamics obtained from Eq.~(\ref{eqn:ALMasterEq}) (green solid line) and the axial motion model of Eq.~(\ref{eqn:MotionMasterEq}) (red solid line). (b) Single-body echo. Blue markers indicate experimental data for the measured inversion after an echo sequence of duration $t_{echo}$ for $\Omega = 0.94(5)\chi N$. Red solid line is numerical simulation based on Eq.~(\ref{eqn:MotionMasterEq}). }
\label{fig:SI1}
\end{figure*}

Furthermore, to account for the large number of particles in the experiment, which would be computationally taxing to simulate, we rescale parameters $g \to g\sqrt{N/N_{\mathrm{sim}}}$ and $\Omega_p \to \Omega_p \sqrt{N_{\mathrm{sim}}/N}$, where $N_{\mathrm{sim}}$ is the number of atoms (lattice sites) we simulate. The smaller atom number also motivates us to randomly position the $N_{\mathrm{sim}}$ atoms within the optical lattice, and we average over the obtained simulation results. Additional motional decoherence is included with $\gamma_{el}/(2\pi) = 40~$kHz (see Sec.~\ref{sec:ResidualMotion} of SI) and $\gamma_s/(2\pi) = 7.5$~kHz. Lastly, we also include thermal effects due to the initial temperature of the atomic sample by populating the axial trap levels according to a Boltzmann distribution $P_n \propto e^{-E_n/k_B T}$, whilst the radial width $\sigma_{\mathrm{th}}$ of the thermal cloud is incorporated by randomly sampling the radial position $r$ of the atoms from a Gaussian distribution $P(r) \propto e^{-r^2/(2\sigma_{\mathrm{th}}^2)}$ and rescaling the peak atom-light coupling by $g(r) = g e^{-(r/w)^2}$ with $w$ the waist of the cavity mode.

As an example to illustrate the importance of including axial motion, we plot a comparison of the mean-field dynamics obtained from the pseudospin-$1/2$ model of Eq.~(\ref{eqn:ALMasterEq}), the axial motion model of Eq.~(\ref{eqn:MotionMasterEq}) and the observed experimental data for $\Omega/(\chi N) = 0.70(2)$ [panel (iv) of Fig.~2a) in the main text] in Fig.~\ref{fig:SI1}a). We observe that the simpler pseudospin-$1/2$ model correctly predicts the dynamics up to the first oscillation, but does not replicate the decay of the experimental observed oscillations thereafter. On the other hand, including the axial motion replicates the observed decay.

We further certify the role of level-changing processes in the experimental system via a single-particle echo experiment. Specifically, after allowing the system to evolve for some time $t_{\mathrm{echo}}$ we rapidly switch the phase of the external pumping field, $\Omega_p \to -\Omega_p$, effectively quenching the coherent drive $\Omega \to -\Omega$. Operating at large $\vert \Omega_p \vert$ deep in the paramagnetic phase, means that the effect of spin-spin interactions mediated by the cavity is negligible for the dynamics. In principle then, in the absence of other single-particle effects the dynamics due to the driving field should be reversed and the initial inversion should be restored after a second evolution of duration $t_{\mathrm{echo}}$. However, the single-particle term describing the trapping potential in Eq.~(\ref{eqn:HAmotion}) is unaffected by the field quench, and results in an imperfect revival of collective atomic observables to their initial conditions at the conclusion of the echo protocol. This is illustrated in Fig.~\ref{fig:SI1}b), wherein we plot experimental data of a single-body echo with the drive chosen to lie deep in the paramagnetic phase, $\Omega/(\chi N) = 0.94(5)$, which corresponds to $\Omega/(2\pi) = 2.28(8)~$MHz, much bigger than the typical axial trap frequency around 200~kHz. We observe the revival of the inversion to the initial value along the south pole, $\langle \hat{J}_z \rangle/(N/2) = -1$, rapidly degrades as the total echo time $t_\textrm{echo}$ is increased. This is due to the presence of the non-commuting trap term  Eq.~(\ref{eqn:HAmotion}), which becomes relevant because the occupation of each trap level is not conserved, i.e. the strong drive leads to Rabi flopping on the optical transition which is accompanied by level-changing processes as described by Eq.~(\ref{eqn:HAL_motion}). A numerical solution of Eq.~(\ref{eqn:MotionMasterEq}) within the mean-field approximation shows excellent agreement, validating our interpretation. In our experiment, these effects start to be relevant for drives $\Omega \gtrsim 0.25 \chi N$.

\section{Anomalous decoherence due to residual motion and technical noise \label{sec:ResidualMotion}}
While we have described in the previous section a minimal addition to our atom-light model to incorporate some effects of axial motion, there still exists residual decoherence due to, e.g., radial motion in the experimental system. As previously foreshadowed, we crudely incorporate this into our model by an additional single-particle homogeneous broadening term in the master equation [Eq.~(\ref{eqn:ALMasterEq})].

\begin{figure*}[htb!]
\includegraphics[width = 6.75in ]{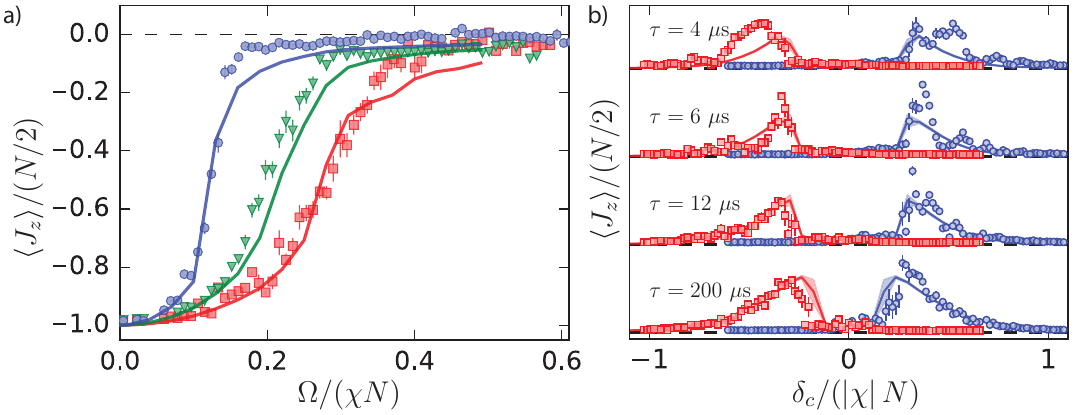}
\caption{{\bf Critical point and decay of coherence.} (a) We study the two dynamical  phases as function of the drive amplitude for different evolution times $\tau$. Red, green and blue markers indicate the atomic magnetization $\langle \hat{J}_z(\tau) \rangle$ for $\tau = 4,~12,~200 ~\mu$s respectively. Solid traces represent simulated dynamics under the described decoherence effects for the same $\tau$. The critical point shift towards lower values of $\Omega$ is consistent with our simplified model. (b) Magnetization $\langle \hat{J}_z(\tau) \rangle$ after having evolved for $\tau = 4,~6,~12, ~200~\mu$s, as indicated above each trace, as a function of the drive detuning $\delta$ with respect to the atomic transition. Drive amplitude is $\Omega = 0.070(5)\chi N$, below the experimentally observed critical point $\Omega^{\mathrm{exp}}_c$. Traces are displaced vertically, but they are not re-scaled. Horizontal dashed black lines indicate $\langle \hat{J}_z \rangle/(N/2) = -1$ for each trace. }
\label{fig:SI2}
\end{figure*}

We empirically confirm the form of this decoherence by analysis of the experimental data, in particular comparing the measured inversion $\overline{\langle \hat{J}_z \rangle}$ across a range of time scales. In Fig.~\ref{fig:SI2}a) we drive the system for some duration $\tau$ and record the inversion at that final time for different drive strengths $\Omega$. For short times ($\tau = 4~\mu$s), red markers, we observe the transition around $\Omega \sim 0.35 \chi N$. As time increases, green markers for $\tau = 12~\mu$s and blue markers for $\tau = 200~\mu$s, the critical point moves towards lower values of $\Omega$, consistent with the model predictions. The reduction in the critical drive power can be explained by the increased loss of collective coherence for longer times, which leads to a degradation in the influence of the global interactions. This is simplest to understand within the paradigm of the collective model which we use for intuition, wherein single-particle broadening would break the collective symmetry and reduce the collective enhancement $\propto N$ of the interaction term. 

Further evidence to support the single-particle broadening is given by analyzing the response of the inversion to the detuning $\delta$ of the coherent drive, in the spirit of Fig.~3 of the main text. In Fig.~\ref{fig:SI2}b) we plot the inversion $\overline{\langle \hat{J}_z \rangle}$ as a function of detuning $\delta$ in the ($\delta=0$) superradiant phase, $\Omega = 0.070(5)\chi N$. We present data where measurement of the inversion is made after different evolution times $t=(4,6,12,200)~\mu s$. The magnitude of the critical $\delta_c$, at which the dynamics switches between the two dynamical phases, is reduced as the evolution time is increased. The reduction is again consistent with the interpretation of Fig.~\ref{fig:SI2}a), i.e., due to the destruction of collective coherence by single-particle broadening. As in Fig.~3 of the main text, we present data for different cavity detunings $\Delta/(2\pi) = \pm 50$~MHz. This allows us to demonstrate that the response of the dynamics is symmetric and the observed shifts are not attributable to other single-particle effects, e.g., residual shifts of the probe detuning $\delta$.

\section{Dynamical phase diagram}
In the main text we presented a dynamical phase diagram [Figs.~1(b) and (c)] for the collective XY model with transverse and longitudinal fields in the mean-field limit. Here, we discuss how this phase diagram was calculated and derive an equation for the critical point separating the dynamical phases.

To be concrete, we consider the collective limit of Eq.~(\ref{eqn:InhomHamiltonian}) derived in Sec.~\ref{sec:SpinModel} of this SI, so the Hamiltonian reduces to:
\begin{equation}
    \hat{H}= \chi\hat{J}^+\hat{J}^-+\Omega\hat{J}_x-\delta\hat{J}_z,
\end{equation}
where $\hat{J}_{\alpha}=\sum_j\sigma^{\alpha}_j/2$ are collective spin operators for $\alpha = x,y,z$ with $\hat{J}^{\pm} = \hat{J}_x \pm i\hat{J}_y$ and we have set $\hbar=1$ herein. The order parameter for the DPT is rigorously defined as the infinite time average of the magnetization:
\begin{equation}
    \overline{\braket{\hat{J}_z}}=\lim_{T\to\infty}\frac{1}{T}\int_{0}^T\braket{\hat{J}_z(t)}\,dt .
\end{equation}
For concreteness, we consider the situation where all atoms start in their ground state at $t=0$, i.e. $\braket{\hat{J}_z(0)}=-N/2$.

The behaviour of the time-averaged magnetization  $\overline{\braket{\hat{J}_z}}$ as a function of $\Omega$, $\chi$ and $\delta$ can be exactly determined in the mean-field limit. We first calculate the equation of motion for $\hat{J}_z$:
\begin{equation}
    \dot{\hat{J}}_z=\Omega\hat{J}_y .
\end{equation}
Within the mean field approximation we treat the quantum operators $\hat{J}_{\alpha}$ as c-numbers and consider the re-scaled variables $s_{\alpha}=\hat{J}_{\alpha}/(N/2)$. The equation of motion is then supplemented with two conservation laws:
\begin{equation}
    s_x^2+s_y^2+s_z^2=1,
\end{equation}
and
\begin{equation}
    \frac{\chi N}{2}(s_x^2+s_y^2)-\delta s_z+\Omega s_x=\frac{E}{N/2},
\end{equation}
which correspond to conservation of the spin length and energy $E$ respectively. For the latter we have $E/(N/2)=\delta$ as a consequence of our specific initial conditions. These additional constraints allow us to solve for $s_y$ in terms of $s_z$ so that in the end the time evolution of $s_z$ reduces to a single differential equation:
\begin{equation}
    (\dot{s_z})^2+V(s_z)=0,
\end{equation}
where
\begin{equation}
    V(s_z)=\Omega^2(1+s_z)\Bigg\{(1+s_z)\bigg[\frac{\chi N s_z}{2\Omega}-\frac{(\chi N/2-\delta)}{\Omega}\bigg]^2+s_z-1\Bigg\}.
\end{equation}
The last two equations are analogous to the equation of motion of a particle with coordinate $s_z$ moving in one dimension and subject to the potential $V(s_z)$. The traditional analysis of such a situation shows that the qualitative behaviour of the system is governed by the zeroes of $V(s_z)$.

\begin{figure}
    \centering
    \includegraphics[width=0.48\textwidth]{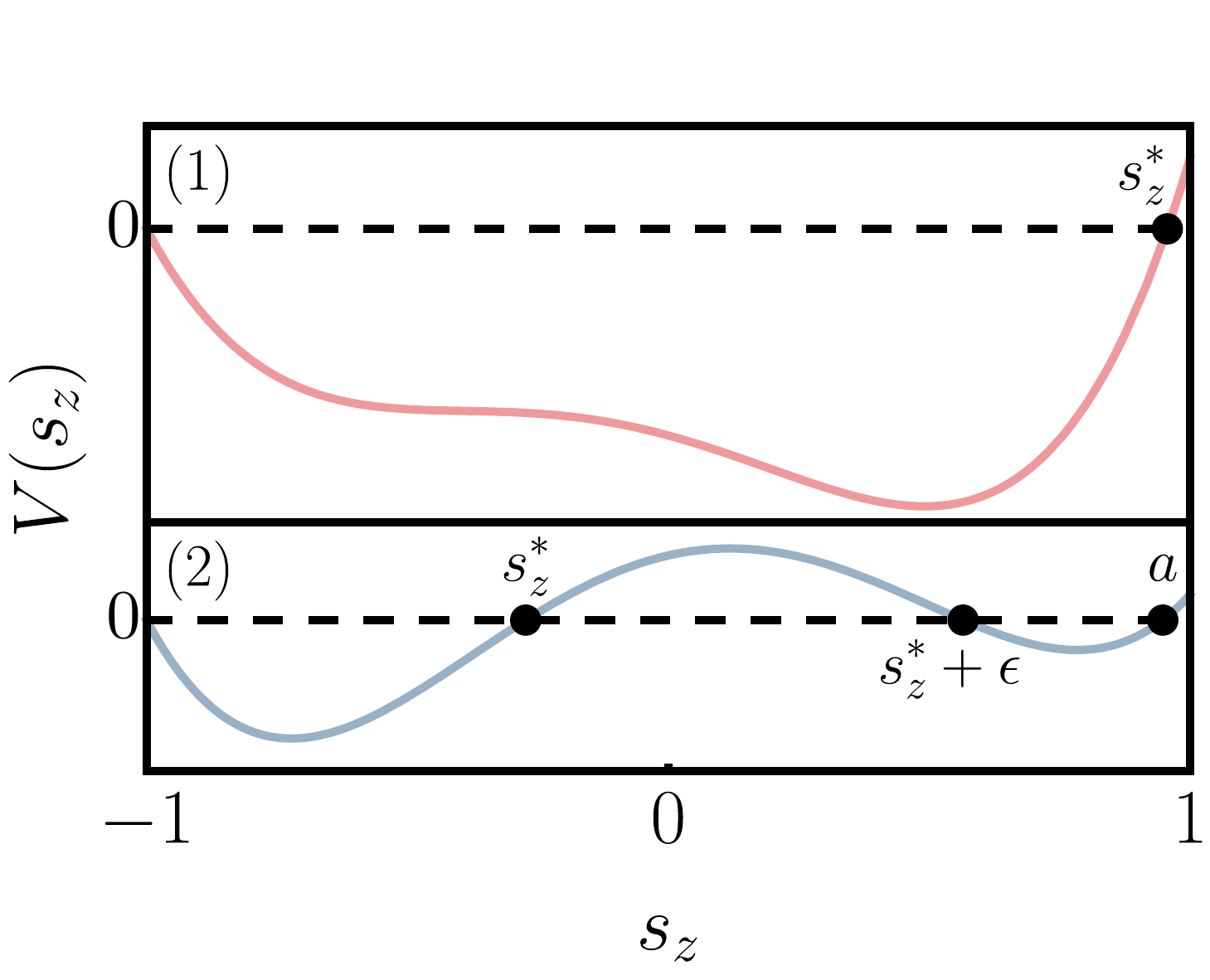}
    \caption{\textbf{Qualitative behaviour of $V(s_z)$}. Depending on the values of $\Omega$, $\delta$ and $\chi$, the effective potential $V(s_z)$ describing the time evolution of $s_z(t)$ shows two different forms: (1) only two real zeroes (upper curve) or (2) four real zeroes (lower curve) in the range $s_z\in[-1,1]$. The non-analyticity in the long time average of $s_z(t)$ appears when $V$ switches between these two situations.}
    \label{fig:SI_EffectivePotential}
\end{figure}

As shown in Fig.~\ref{fig:SI_EffectivePotential}, there are two qualitatively different forms $V(s_z)$ can take: (1) there are two real zeroes, one at $-1$ and a second one that we denote as $s_z^*$, and (2) $V(s_z)$ has four real zeroes, which we denote as $-1$, $s_z^*$, $s_z^*+\epsilon$ and $a$ in increasing order. In both scenarios, $s_z(t)$ has the value $-1$ at $t=0$, but will increase with time until it reaches $s_z^*$. After that, $s_z(t)$ will return to $-1$ and repeat this motion in a periodic fashion. As we will show below, the infinite time average of $s_z(t)$ will develop a non-analyticity precisely when $V(s_z)$ switches from having $4$ to having only $2$ real zeroes. This occurs when
\begin{align}
    \begin{split}
        V(s_z)&=0\\
        V'(s_z)&=0
    \end{split}
\end{align}
which gives a relation between $\Omega$ and $\delta$:
\begin{align}\begin{split}\label{eqn:SICriticalOmega}
    \frac{\Omega_c(\delta)}{\chi N}=\frac{1}{2}&\Bigg[2\bigg(1-\frac{\delta}{\chi N}\bigg)\bigg(1+\frac{2\delta}{\chi N}\bigg)\\&-\frac{3}{2}\bigg(\frac{8\delta}{\chi N}+1\bigg)+\frac{1}{2}\bigg(1+\frac{8\delta}{\chi N}\bigg)^{3/2}\Bigg]^{1/2}.
\end{split}\end{align}
Furthermore, we can also calculate the form of the non-analyticity in the time average. Close to the transition, and in the region with four real zeroes, $V(s_z)$ has the form:
\begin{equation}
    V(s_z)=\frac{\chi^2 N^2}{4}(1+s_z)(s_z-s_z^*)(s_z-s_z^*-\epsilon)(s_z-a),
\end{equation}
where the zeroes, in ascending order, are $-1$, $s_z^*$, $s_z^*+\epsilon$ and $a$. For a fixed $\delta$, when $\Omega\to\Omega_c(\delta)$, $\epsilon\to 0$ as a power of $|\Omega-\Omega_c(\delta)|$. Then,
\begin{equation}
    \int_{0}^Ts_z(t)\,dt=2\int_0^{s_z^*}\frac{s_z\,ds_z}{\sqrt{-V(s_z)}}\sim (s_z^*)^c A\log(\epsilon)+B,
\end{equation}
where $A$ and $B$ are finite constants, $(s_z^*)^c$ is $s_z^*$ evaluated at $\Omega=\Omega_c(\delta)$ and we have neglected terms that go to $0$ as $\epsilon\to 0$. In the same way
\begin{equation}
    T=\int_0^Tdt=2\int_0^{s_z^*}\frac{ds_z}{\sqrt{-V(s_z)}}\sim A\log(\epsilon)+B',
\end{equation}
where $B'$ is another constant. Hence, close to the transition
\begin{equation}
   \frac{ \overline{\braket{\hat{J}_z}}}{N/2}\to (s_z^*)^c+\frac{C}{\log{\epsilon}}=(s_z^*)^c+\frac{C'}{\log{\Big|\Omega_c(\delta)-\Omega\Big|}},
\end{equation}
where $C$ and $C'$ are yet other constants. We have thus shown that the non-analyticity is logarithmic. A similar analysis on the other side of the transition shows the same logarithmic behaviour (albeit with a different constant $C'$). For the sake of completeness, we include the expression for $(s_z^*)^c$ as a function of $\delta$:
\begin{equation}
    (s_z^*)^c=\frac{1}{2}-\frac{1}{2}\sqrt{1+\frac{8\delta}{\chi N}}.
\end{equation}
For Eq.~(\ref{eqn:SICriticalOmega}) to make sense, the condition $\delta>-\chi N/8$ needs to be satisfied. When this inequality is violated, then $V(s_z)$ has only two real zeroes for any value of $\Omega$. Hence there is no transition and only a smooth crossover.

\section{Mapping between spin model and macroscopic self-trapping \label{sec:SelfTrapping}}

The connection between the collective spin model, Eq.~(1) of the manuscript, and the phenomena of macroscopic self-trapping and Josephson oscillations observed in atomic and solid-state polariton condensates can be made by an appropriate mapping between the spin and bosonic models.  

For the condensate examples a simple description of self-trapping and Josephson oscillations can be obtained by a two-well model in the single-mode approximation \cite{Diaz_2010} with the Hamiltonian given by
\begin{multline}
    \hat{H} = \frac{U}{2}\left[ \hat{n}_R(\hat{n_R}-1) + \hat{n}_L(\hat{n_L}-1) \right] \\ + \mathcal{J}\left(\hat{a}^{\dagger}_R\hat{a}_L + \hat{a}^{\dagger}_L\hat{a}_R \right) + \frac{\epsilon}{2}\left(\hat{n}_L - \hat{n}_R \right) \label{eqn:TwoWellTrapping}
\end{multline}
where $\hat{a}_{R(L)}$ is the bosonic annihilation operator of the right (left) well and $\hat{n}_{R(L)} = \hat{a}^{\dagger}_{R(L)}\hat{a}_{R(L)}$. Contact interactions between the constituent atoms of the condensate are characterized by $U$, and $\mathcal{J}$ defines the tunnelling between the two wells and $\epsilon$ is some controllable energy difference. 

Using that total particle number $\hat{n}_R + \hat{n}_L$ is conserved, and introducing the Schwinger boson mapping $\hat{J}_z = (1/2)(\hat{n}_L - \hat{n}_R)$, $\hat{J}^+ = \hat{a}^{\dagger}_R\hat{a}_L$ and $\hat{J}^- = \hat{a}^{\dagger}_L\hat{a}_R$, we can recover from Eq.~(\ref{eqn:TwoWellTrapping}):
\begin{equation}
    \hat{H} = U\hat{J}_z^2 + 2\mathcal{J}\hat{J}_x + \epsilon\hat{J}_z + \frac{U}{4} N(N-2). \label{eqn:TwoWellSz}
\end{equation} Note that the total number of particles $N=(\hat{n}_L + \hat{n}_R)$ commutes with the Hamiltonian and therefore is a conserved quantity.
Finally, using that $\hat{J}^2$ is also a conserved quantity of this Hamiltonian, and that $\hat{J}_z^2 \equiv \hat{J}^2 - \hat{J}^+\hat{J}^- + \hat{J}_z$, we have that Eq.~(\ref{eqn:TwoWellSz}) is identical to the collective XY model with transverse and longitudinal fields upon recognizing: $\chi = -U$, $\Omega = 2\mathcal{J}$ and $\delta = U + \epsilon$.


\end{document}